\documentclass[traditabstract]{aa}
\usepackage{graphicx}
\usepackage{txfonts}
\usepackage{natbib}
\usepackage{enumitem}
\bibpunct{(}{)}{;}{a}{}{,}

 \newcommand{\mics}{$\mu$m}
 \newcommand{\lums}{$\nu L _\nu (60\mu $m$)$}
 \newcommand{\lfir}{$L _{60}$}
 \newcommand{\lfirm}{$L^{b} _{60}$}
 \newcommand{\lagn}{$L _{AGN}$}
 \newcommand{\lagnc}{$L ^{c}_{AGN}$}
 \newcommand{\lhard}{$L _{X}$}
 \newcommand{\ergs}{erg s$^{-1}$}
 \newcommand{\kms}{km s$^{-1}$}
 \newcommand{\mmir}{$m_{24\textrm{-}160}$}
 \newcommand{\mfir}{$m_{100\textrm{-}160}$}

\begin{document}

\title{The Mean Star Formation Rate of X-ray selected Active Galaxies and its Evolution from $z\sim2.5$: Results from PEP-Herschel \thanks{{\it Herschel} is an ESA space observatory with science instruments provided by European-led Principal Investigator consortia and with important participation from NASA.}}

\author{
D. J. Rosario\inst{1}
\and
P. Santini\inst{1,2}
\and
D. Lutz\inst{1}
\and
L. Shao\inst{1}
\and
R. Maiolino\inst{2}
\and
D. M. Alexander\inst{3}
\and
B. Altieri\inst{4}
\and
P. Andreani\inst{5,6}
\and
H. Aussel\inst{7}
\and
F. E. Bauer\inst{8,9}
\and
S. Berta\inst{1}
\and
A. Bongiovanni\inst{9,10}
\and
W. N. Brandt\inst{11}
\and
M. Brusa\inst{1}
\and
J. Cepa\inst{10,12}
\and
A. Cimatti\inst{13}
\and
T. J. Cox\inst{14}
\and
E. Daddi\inst{7}
\and
D. Elbaz\inst{7}
\and
A. Fontana\inst{2}
\and
N. M. F{\"o}rster Schreiber\inst{1}
\and
R. Genzel\inst{1}
\and
A. Grazian\inst{2}
\and
E. Le Floch\inst{7}
\and
B. Magnelli\inst{1}
\and
V. Mainieri\inst{5}
\and
H. Netzer\inst{15}
\and
R. Nordon\inst{1}
\and
I. P{\'e}rez Garcia\inst{10,12}
\and
A. Poglitsch\inst{1}
\and
P. Popesso\inst{1}
\and
F. Pozzi\inst{12}
\and
L. Riguccini\inst{7}
\and
G. Rodighiero\inst{16}
\and
M. Salvato\inst{17,1}
\and
M. Sanchez-Portal\inst{4}
\and
E. Sturm\inst{1}
\and
L. J. Tacconi\inst{1}
\and
I. Valtchanov\inst{4}
\and
S. Wuyts\inst{1}
}

\offprints{D.J. Rosario \email{rosario@mpe.mpg.de}}

\institute{ Max-Planck-Institut f\"{u}r Extraterrestrische Physik (MPE), Postfach 1312, 85741 Garching, Germany.
\and INAF - Osservatorio Astronomico di Roma, via di Frascati 33, 00040 Monte Porzio Catone, Italy.
\and Department of Physics, Durham University, South Road, Durham, DH1 3LE, UK.
\and European Space Astronomy Centre, Villafranca del Castillo, Spain.
\and ESO, Karl-Schwarzschild-Str. 2, D-85748 Garching, Germany.
\and INAF-Osservatorio Astronomico di Trieste, via Tiepolo 11, 34131 Trieste, Italy.
\and Laboratoire AIM, CEA/DSM-CNRS-Universit{\'e} Paris Diderot, IRFU/Service d'Astrophysique, B\^at.709, CEA-Saclay, 91191 Gif-sur-Yvette Cedex, France.
\and Pontificia Universidad Cat\'olica de Chile, Departamento de Astronom\'ia y Astrof\'isica, Casilla 306, Santiago 22, Chile.
\and Space Science Institute, 4750 Walnut Street, Suite 205, Boulder, CO 80301, USA
\and Instituto de Astrof{\'i}sica de Canarias, 38205 La Laguna, Spain.
\and Department of Astronomy and Astrophysics, 525 Davey Lab, Pennsylvania State University, University Park, PA 16802, USA.
\and Departamento de Astrof{\'i}sica, Universidad de La Laguna, Spain.
\and Dipartimento di Astronomia, Universit{\`a} di Bologna, Via Ranzani 1, 40127 Bologna, Italy.
\and Carnegie Observatories, 813 Santa Barbara Street, Pasadena, CA, 91101, USA
\and School of Physics and Astronomy, Tel Aviv University, 69978 Tel Aviv, Israel.
\and Dipartimento di Astronomia, Universit{\`a} di Padova, Vicolo dell'Osservatorio 3, 35122 Padova, Italy.
\and IPP-Max-Planck-Institut f\"{u}r Plasmaphysik, Boltzmannstrasse 2, D-85748, Garching, Germany.
}

\date{Received .... ; accepted ....}
\titlerunning{SFR of X-ray selected Active Galaxies}


 \abstract{  
 
 We study relationships between star-formation rate (SFR) and the accretion luminosity and nuclear obscuration
 of X-ray selected Active Galactic Nuclei (AGNs) using a combination of deep far-infrared (FIR) and X-ray data
 in three key extragalactic survey fields (GOODS-South, GOODS-North and COSMOS), as part of the
 PACS Evolutionary Probe (PEP) program. The use of three fields with differing areas and depths enables
 us to explore trends between the global FIR luminosity of the AGN hosts and the luminosity of the active
 nucleus across 4.5 orders of magnitude in AGN luminosity (\lagn) and spanning redshifts from the Local Universe
 to $z=2.5$. Using imaging from the Herschel/PACS instrument in 2-3 bands, we combine FIR detections and
 stacks of undetected objects to arrive at mean fluxes for subsamples in bins of redshift and X-ray luminosity.
 We constrain the importance of AGN-heated dust emission in the FIR and confirm that the majority of the FIR 
 emission of AGNs is produced by cold dust heated by star-formation in their host galaxies.
 
 We uncover characteristic trends between the mean FIR luminosity (\lfir) and accretion luminosity of
 AGNs, which depend both on \lagn\ and redshift. At low AGN luminosities, accretion and SFR are
 uncorrelated at all redshifts, consistent with a scenario where most low-luminosity AGNs are primarily
 fueled by secular processes in their host galaxies. At high AGN luminosities, a significant correlation is
 observed between \lfir\ and \lagn, but only among AGNs at low and moderate redshifts ($z<1$). 
 We interpret this observation as a sign of the increasing importance of major-mergers in driving both the
 growth of super-massive black holes (SMBHs) and global star-formation in their hosts
 at high AGN luminosities.  We also find evidence that the enhancement of SFR in luminous AGNs 
 weakens or disappears at high redshifts ($z>1$) suggesting
 that the role of mergers is less important at these epochs. 
 
 At all redshifts, we find essentially no
 relationship between \lfir\ and nuclear obscuration across five orders of magnitude in obscuring
 Hydrogen column density ($N_H$), suggesting that various
 mechanisms are likely to be responsible for obscuring X-rays in active galaxies.
 
 We discuss a broad scenario which can account for these trends: one in which two different modes of 
 AGN fueling operate in the low- and high-luminosity regimes of SMBH accretion. We postulate
 that the dominant mode of accretion among high-luminosity AGNs evolves with redshift. Our study, as well as a 
 body of evidence from the literature and emerging knowledge about the properties of high redshift galaxies, 
 supports this scenario.   
  }

\maketitle

\section{Introduction}

 A number of different lines of evidence suggest that a close relationship
 exists between the growth of supermassive black holes (SMBHs) and their
 host galaxies. There are now well-established and tight correlations between the properties of nuclear 
supermassive black holes (SMBHs) and the spheroids of their host galaxies 
\citep{magorrian98, ferrarese00, ferrarese01, marconi03, ferrarese05, gultekin09, gebhardt00, tremaine02}.
The strong evolution in the cosmic star-formation rate (SFR) density \citep[e.g.,][]{hopkins06}
is also reflected in the evolution of the number density of bright Quasi-stellar Objects (QSOs)
and the cosmic SMBH accretion rate density since $z\sim3$ \citep{boyle98, ueda03, croom04, silverman08,
aird10, assef11}. In addition, AGNs, like star-forming galaxies \citep{fontanot09},
display a form of  `down-sizing', by which the population that dominates either nuclear or star-forming
activity moves to systems at lower luminosities or masses towards later epochs \citep{cowie03, fiore03,
hasinger05, bongiorno07}. 

Observational studies of the relationship between host 
star-formation and AGN activity are an essential test of the degree 
and importance of galaxy-SMBH co-evolution scenarios.
Traditionally, such studies of star-formation in AGNs have concentrated on interesting subsamples, usually
luminous or rare types, such as bright QSOs, radio-loud AGNs or bright local Seyfert galaxies. 
Many of these types of AGNs are associated with on-going starbursts \citep{rowanrob95, haas98,
omont03, priddey03, jahnke04, shi05, schweitzer06, netzer07, greene09}. In general, one finds clear star-formation
signatures in a sizable fraction of bright QSOs (as much as 80\%, depending on selection), 
with some reaching infra-red (IR) luminosities placing them in the 
ultra-luminous IR galaxy (ULIRG) regime \citep[e.g.,][]{page04, stevens05, alexander08, coppin08}. 
In a similar vein, a significant fraction of ULIRGs/sub-mm galaxies
show clear AGN signatures, though not always at the level of QSOs \citep{sanders88, sanders96, genzel98,
canalizo01}. These studies support the notion of a connection between galaxy mergers and black hole growth, 
but, by themselves, they do not explore star-formation in the bulk of AGNs.

With the advent of large systematic galaxy surveys, several studies have turned to complete, well-defined
samples of AGNs selected by multiple means and covering a large dynamic range in AGN 
luminosity and obscuration. Star formation in these AGN hosts is revealed
typically through photometric or spectroscopic
signatures at optical or UV wavelengths. \cite{kauffmann03} calibrated the star-formation history
of emission-line selected AGNs in the local Universe using spectra from Sloan Digital Sky Survey (SDSS)
and found that the mean SFR and typical ages of AGN host galaxies are respectively higher and younger than
those of inactive galaxies. Post-starburst spectroscopic signatures are also found to be strong
in AGN hosts \citep{wild07}, while there is evidence in some AGNs that nuclear activity follows 
a few 100 Myr after a strong starburst \citep{davies07, wild10}.

At higher redshifts, multi-wavelength surveys such as GOODS \citep{dickinson03}, COSMOS \citep{scoville07}
and AEGIS \citep{davis07} played a major role.
\cite{silverman09}, using the SF sensitive [O II]$\lambda 3727$ emission line, 
found wide-spread SF in low and moderate luminosity AGN hosts at levels comparable to 
inactive galaxies of similar stellar mass. AGN host colors paint a similar picture
\citep{xue10,rosario12}. Clustering studies suggest that AGNs 
are associated more with red or evolved galaxies, rather than the bulk of
star-forming galaxies \citep{coil09}. These studies generally imply that low-luminosity AGNs are hosted by
a special category of galaxy that is quite evolved but still forming stars. Since SFR correlates quite
strongly with stellar mass \citep{noeske07, daddi07, santini09} and AGN hosts are preferentially hosted
by massive galaxies \citep{alonzo08}, stellar mass selection effects can bias any 
inferences about SF in AGN hosts \citep{silverman09, xue10, rosario12}.
When taken into account, active and inactive galaxies are found to be quite similar in their 
average SF properties.

\begin{figure*}[t]
\includegraphics[width=\textwidth]{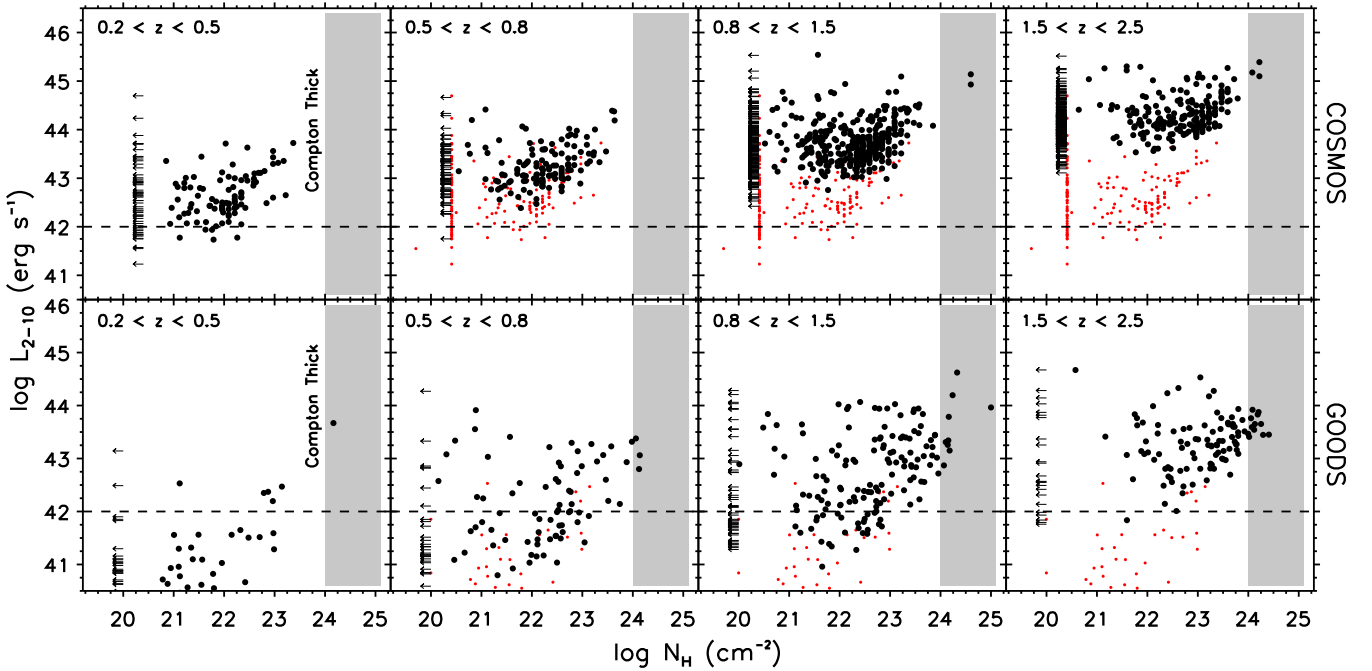}
\caption{Rest-frame 2-10 keV (hard band) X-ray luminosity \lhard\ vs. hydrogen obscuring 
column $N_H$ for X-ray sources in this study.
The upper four panels shows sources in COSMOS, while the lower panels show sources from both GOODS fields
combined. Each panel is for a different bin in redshift, increasing left to right. In the three higher redshift bins, the
location of the X-ray sources from the lowest ($0.2<z<0.5$) bin are shown for reference as small red points, 
to clearly bring out the redshift trends in the sample. Sources with $N_H$ consistent with only Galactic obscuration 
are shown as left-pointing arrows. The fiducial limit of \lhard$=10^{42}$ \ergs\
used to define a clean AGN sample (employed in \S4.3) is shown by a dashed line. The nominal 
Compton thick regime of X-ray obscuration (N$_H > 10^{24}$ cm$^{-2}$) is shown as a shaded region in each panel. 
}
\label{lx_vs_nh}
\end{figure*}

A caveat of most studies of star-formation in AGN hosts is that they tend to concentrate on low and
moderate luminosity AGNs. This is because the emission of the active nucleus can be a strong 
contaminant of SF tracers at
optical and UV wavelengths. Together with the difficulties
stemming from differential dust obscuration and reddening between nuclear and galaxy emission,
this makes it next to impossible to accurately determine SF properties of luminous AGNs using these data.
Tracers in the IR are useful since extinction is rarely a major issue at these
wavelengths (except around the 10 \mics\ Si absorption feature). 
The IR continuum comes primarily from reprocessed short wavelength radiation and
the total IR luminosity (generally integrated over the range of 8-1000 \mics) is considered to be a
good calorimeter of the total bolometric output of both AGNs and SF systems. Since the mid-infrared (MIR) 
has traditionally been more accessible to observation from early generations of ground and space-based instruments, 
several studies of SF in AGNs have been based on the measurement of the prominent PAH emission
features that are excited in star-forming environments \citep{netzer07, lutz08, shi09}. QSOs frequently
show stronger PAH features than typical inactive galaxies, though evolution in the properties of the
massive galaxy population was generally not taken into account in these studies. Another limitation
is that MIR spectroscopy, on which PAH measurements depend, is only sensitive enough
for rather IR-bright galaxies, which are not representative of the broad population
of AGN hosts.

Wide area galaxy surveys from the new Herschel Space Observatory \citep{pilbratt10} 
have opened up the ability to carefully measure the rest-frame far infrared (FIR) continuum
luminosity of distant galaxies and AGNs. Several studies show that the FIR continuum 
is the best and least contaminated tracer of SFR in massive galaxies, which are dusty 
enough for most SF luminosity to be reprocessed into the FIR. In a first study of X-ray
selected AGNs in the GOODS-N field, \cite{shao10} found that the mean relationship between the
\lagn\ and global host galaxy SF is a function of the accretion luminosity, such that
luminous AGNs
show a SF-\lagn\ connection that is absent in lower luminosity systems.
Using a similar approach, \cite{mullaney11b}, using deeper FIR imaging in both GOODS fields,
find that the mean SFRs of X-ray sources is indistinguishable from those of typical inactive star-forming 
galaxies, though the study only concentrated on low to moderate luminosity AGNs. Alternatively,
\cite{hatzimina10} find that FIR luminosity is correlated with AGN luminosity over many orders
of magnitude in both quantities, at apparent odds with aspects of both previous results.

Many of the differences between these works can be reconciled by studying the FIR properties
of a broad sample of AGNs across many epochs. 
We present the largest study to-date of the mean SF properties of AGNs, in terms of the
range of accretion luminosity (5.5 orders of magnitude), SFR and redshift (Local to $z=2.5$). 
This is the second paper of a series
that combines deep FIR images and the best current X-ray analysis
in multiple deep fields to determine the SF characteristics of a large sample of AGNs spanning the redshifts $0.2<z<2.5$. 
Using techniques of FIR stacking, we have constrained the mean FIR luminosities
for active galaxies across redshift, AGN luminosity, obscuration and host mass. 
In the first paper of the series \citep{santini11}, we studied the mean SFR of
AGN hosts as function of stellar mass and redshift, finding a slight enhancement
compared to inactive galaxies with the same stellar mass distribution.
In this work, we build on early work from our team \citep{shao10} and
uncover trends with AGN luminosity and obscuration. The paper is organised as follows:
\S2 describes the datasets used and outlines our methods (which are detailed
in the Appendix), \S3 deals with
AGN emission in the FIR as a source of contamination, \S4 outlines our main trends, fits
and tests for biases. In \S5, we discuss our results in the context of galaxy evolution.

We assume a standard $\Lambda$-CDM Concordance cosmology, with $H_{0} = 70$ \kms~Mpc$^{-1}$.

\section{Datasets, Sample Selection and Methods}

This work exploits deep X-ray and FIR imaging, as well as extensive ancillary data, in three well-studied extragalactic fields: GOODS-South, GOODS-North and COSMOS. The depth of the two GOODS fields in the X-ray, FIR and several other bands is essential to probe faint and high-redshift AGNs. The shallower, wide-area COSMOS field provides us with good statistics among the more luminous sources, which are rare in the smaller GOODS fields. 

\subsection{X-ray catalogs}

Detailed X-ray point source catalogs were compiled for all three fields. These provide rest-frame X-ray luminosities and obscuring columns, source classification, optical associations and redshifts. Where possible, spectroscopic redshifts were used. In the absence of these,
photometric redshifts specifically suited for AGN hosts, based on optimised AGN or hybrid templates, were compiled.

For GOODS-S we used the \cite{luo08} catalog from the 2 Msec Chandra Deep Field-South imaging program. 
A small set of new redshifts from recent spectroscopic efforts \citep{silverman10} were included, and 
photometric redshifts were compiled from the work of \cite{luo10}.

For GOODS-N we used the 2 Msec Chandra Deep Field-North catalog of \cite{alexander03} and an updated version of the classification into AGN and other types of X-ray sources compiled by \cite{bauer04} \citep[see also][]{shao10}.

Finally, for COSMOS, we used the XMM-Newton catalog compiled by \cite{cappelluti09} and the X-ray optical associations and derived properties presented in \cite{brusa10} in its updated version, which includes new redshifts from ongoing (mostly DEIMOS/KECK) spectroscopic campaigns and a few changes in the published redshifts and/or spectroscopic classifications from a re-analysis of the data \citep{mainieri11}. Photometric redshifts are as derived by \cite{salvato09}. 

Rest-frame X-ray luminosities in the hard band (2-10 keV) were estimated 
for all our AGNs, using spectral fits for X-ray
sources with sufficient counts or using scalings based on hardness ratios for faint X-ray sources. Where possible, 
the Hydrogen column densities of obscuring gas towards the X-ray sources ($N_H$) were also estimated. 
The X-ray luminosities were then corrected to account for the level of obscuration, yielding 
absorption-corrected hard-band luminosities (\lhard\ hereafter). It must be kept in mind that 
$N_H$ estimates, especially for obscured AGNs at higher redshifts, can be fairly uncertain.
X-ray luminosities in the hard band, however, are generally more secure, though, of course, 
for heavily obscured sources, the uncertainty in $N_H$ will translate into a greater uncertainty in \lhard.

In this work, we consider the properties of all X-ray sources with \lhard\ greater than $10^{41}$ \ergs. 
At the low-luminosity end (\lhard$<10^{42}$ \ergs), there may be some contamination
from very luminous star-bursts where soft-band emission from star-formation related processes (SNe, high-mass X-ray binaries)
can contaminate the sample. Since we study trends with AGN luminosity in our main analysis, we can easily test for
the importance of such starburst contamination in our sample. In general, we do not find that low luminosity X-ray sources
are more strongly star-forming than moderate luminosity sources, except in the redshift range $0.2-0.5$. We account
for such effects in our analysis.

A sense of the differing depths of the X-ray data in the three fields may be gained from Fig.~ \ref{lx_vs_nh}, which plots \lhard\ against the obscuring column $N_H$ for the X-ray sources from the three catalogs used in this analysis. 
We split the sample into a set of 4 redshift bins: $0.2<z<0.5$, $0.5<z<0.8$,$0.8<z<1.5$ and $1.5<z<2.5$.
The fields are complementary in terms of their X-ray properties: the deep GOODS fields allow a good sampling of faint AGN, while in the 
large-area COSMOS field we probe the rarer and more powerful sources. 

The X-ray sources also display strong trends of $N_H$ with redshift, especially in the COSMOS field. This is governed 
mostly by selection effects \citep{tozzi06}. At all redshifts, a substantial fraction of AGNs
have absorbing columns
consistent with no intrinsic absorption (i.e, only absorption by gas in our own Galaxy). 
Towards higher redshifts, the rest-frame soft X-ray energies redshift out of the
detection bands of both Chandra and XMM-Newton, leading to greatly reduced accuracy in the measurement of $N_H$
in weakly absorbed sources, as well as smaller range of estimated $N_H$ values 
from most spectral fitting methods. This explains the large number of sources clustered around 
log N$_H = 20.5$ cm$^{-2}$ in the upper panels. In addition, absorbed low-luminosity AGNs are badly incomplete
in flux-limited X-ray surveys, leading to an apparent correlation between $N_H$ and \lhard, which can be
seen most clearly in the high redshift panels of Fig.~\ref{lx_vs_nh}. While the span of \lhard\ in any given redshift
bin is fairly broad, the span of $N_H$ that we can probe in a bin can be quite limited. 

\subsection{PACS Imaging and catalogs}

Our FIR data come from observations by the PACS instrument \citep{poglitsch10} on board the Herschel 
Space Observatory, as part of the PACS Evolutionary Probe \citep[PEP,][]{lutz11} survey. 
Observations of GOODS-S were carried out in all three PACS bands (70, 100 and 160 \mics), 
while the other two fields were only observed in the two long wavelength bands. 
We made use of PACS catalogs extracted using prior knowledge of the positions and fluxes of sources 
detected in deep archival IRAC 3.6 \mics\ and MIPS 24 \mics\ imaging in these fields. This allows us to 
deblend PACS sources in images characterized by a large PSF, especially in crowded fields, 
and improve the completeness of faint sources at the detection limit. 
Fluxes in GOODS-S reach 1.1, 1.2 and 2.4 mJy at 3$\sigma$ in the 70, 100 and 160 \mics\ bands respectively. In the other two 
fields the 3$\sigma$ flux limit at 100/160 \mics\ is 3.0/5.7 mJy in GOODS-N and 5.0/10.2 mJy in COSMOS \citep{berta11}.  
Detailed information on the PEP survey, observed fields, data processing and source extraction may be found in \cite{lutz11}.

X-ray sources were matched to PACS sources through their optical counterparts. In most cases, a robust 
crossmatch could be made between the optical counterpart of an X-ray source and a 24 \mics\ source 
in the catalogs used to extract PACS fluxes. In the few cases where the crossmatches were not robust, 
either due to uncertain associations between optical and 24 \mics\ sources, or due to multiplicity in a match, 
we visually examined the matches on the 24 \mics\ images to verify their quality and discard unreliable matches.
The technique used to  associate PACS fluxes  with optical counterparts could in principle introduce 
some biases in the determination of the mean FIR luminosity, since AGNs are typically brighter than inactive
galaxies at MIR wavelengths (e.g, 24 \mics). In \cite{santini11}, we show that this bias is unimportant.

\subsection{Swift BAT sample}

For a comparison to local X-ray selected AGNs, we also compiled an unbiased sample of extremely 
hard X-ray-selected AGN (15-150 keV band) from the 39 month Palermo Swift-BAT catalog \citep{cusumano09}.
This sample has also been used in \cite{lutz10} and \cite{shao10}. 

From the Swift-BAT catalog, we selected sources classified as
Seyferts, LINERs, quasars, and other AGNs. We excluded objects with evidence for a strong non-thermal contribution 
to the far-infrared (on the basis of the NED SED) or those classified as blazars. In addition, objects at Galactic latitude 
$|b| < 15$ and objects at redshift $z > 0.3$ were also removed. For the remaining 293 AGNs
we used the IRAS Faint Source Catalog 60 \mics\ detections where available, otherwise we used 
SCANPI\footnote{http://scanpiops.ipac.caltech.edu/applications/Scanpi/index.html} to obtain 60 \mics\ 
measurements for faint or individually nondetected objects. We calculated rest-frame $2$-$10$ keV luminosities 
extrapolating from the BAT fluxes and the redshift, assuming an AGN photon index of 1.8. The sources were averaged in 7 bins of
L$_X$, with sufficient statistics in each bin. 

\subsection{Methods}

AGNs in all three deep fields were divided into four intervals in 
redshift (0.2-0.5, 0.5-0.8, 0.8-1.5, 1.5-2.5), spanning the epochs where most black hole 
growth in the Universe has occurred. In each redshift interval,
the AGNs were further divided by \lhard\ into five bins (41-42, 42-43, 43-44, 44-45, 45-45.5 in $\log L_X$).
An estimate of the mean FIR luminosity ($\nu L _{\nu}$ at 60 $\mu$m) in each bin was made by
combining PACS photometry for FIR detected AGNs and PACS stacks for FIR-undetected AGNs
in each field, and then appropriately combining measurements from all fields together for a single
measurement per bin (\lfir). Details of the PACS photometry, stacking, error estimation and methods of combination
between and across fields are developed in the Appendix. 

\section{AGN emission in the FIR}

The key results in this work are based on the notion that rest-frame FIR emission is
a proxy of SF activity.  A number of previous studies support our assumption based on different grounds.
Many authors \citep[e.g.][]{schweitzer06,netzer07,lutz08}  showed a strong correlation between FIR 
luminosity and SFR tracers, such as PAH emission features, both in local and high redshift 
bright ($L_{AGN}/L_{FIR}$ up to $\sim 10$) QSOs.
In this section, we explore the degree to which AGN emission may contribute at FIR wavelengths, in particular
at 60 \mics. 

\begin{figure*}[ht]
\includegraphics[width=\textwidth]{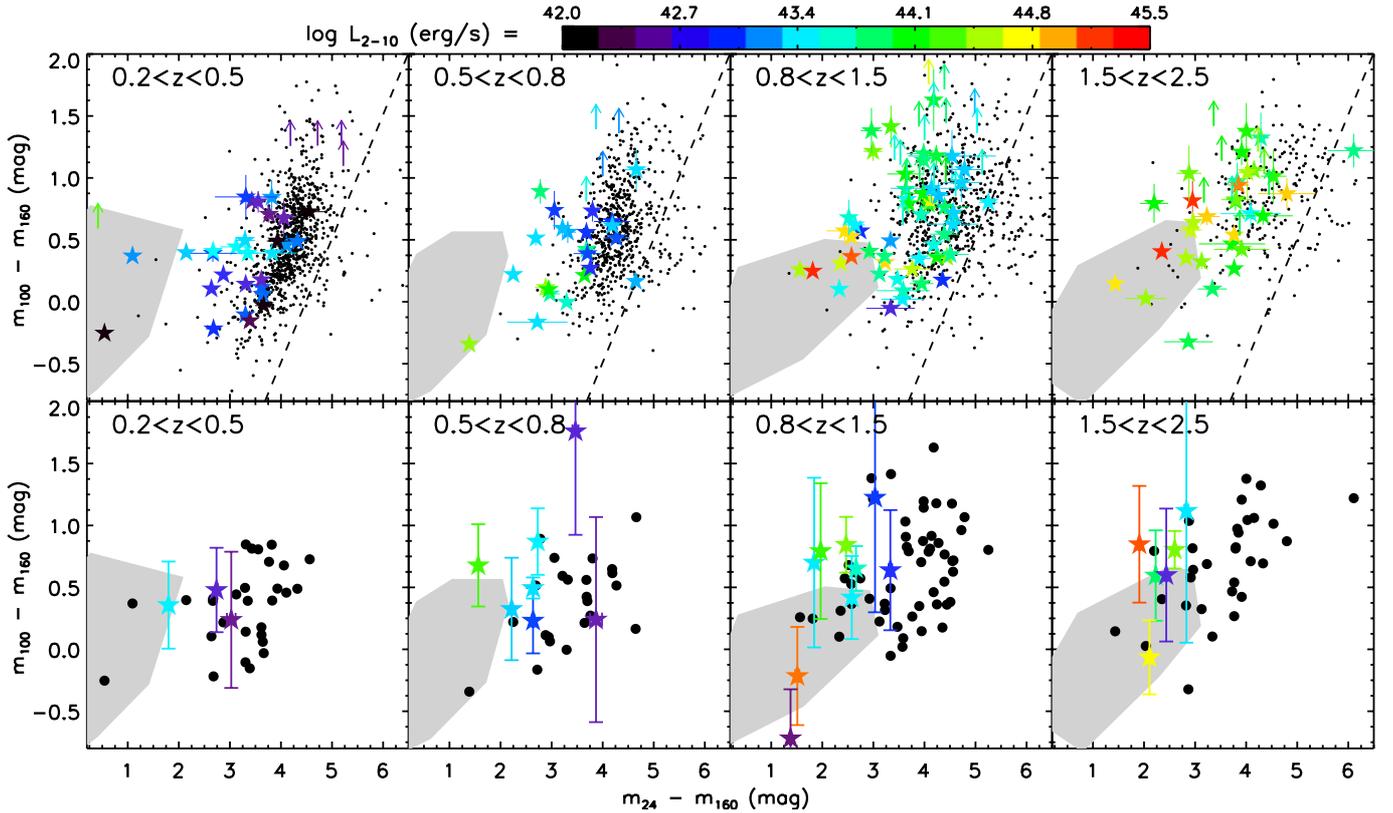}
\caption{Observed IR colors of X-ray AGNs plotted in bins of redshift.
A MIR flux ratio (the flux at 24\mics\ to the flux at 160 \mics, expressed as a magnitude) is
plotted against an FIR flux ratio (the flux at 100 \mics\ to the flux at 160 \mics, also as a magnitude).
The shaded grey regions mark out parts of the diagram
occupied by star-forming galaxies with an AGN fraction of $>50$\% at rest-frame 60 \mics\ (\S3.2). 
{\bf Top row panels:} PACS-detected AGNs from the COSMOS field are shown as star-shaped points, with colors that represent their 
hard-band X-ray luminosity $L_X$  (shown in the color bar at top). The small black dots are all galaxies that are detected in 
the COSMOS PACS maps in the same redshift bins. Objects with colors right and below the dashed line are unlikely 
to be detected because of the photometric limits of the MIR and FIR catalogs.
{\bf  Bottom row panels:} Stacked measurements for PACS-undetected AGNs from all three fields
are shown as star-shaped points, colored by $L_X$. For reference, the large black points 
show the locations of PACS-detected COSMOS AGNs (the same as the colored points in the top panels).
PACS-undetected AGNs have, on average, warmer IR colors and may be more AGN dominated
in the FIR. 
}
\label{cosmos_ccdiag}
\end{figure*}

\subsection{Predictions from Models of AGN-heated dust}

The nature of emission from AGN-heated dust has been the topic of several theoretical studies
\citep{pier92, granato94, efstathiou95, nenkova02, fritz06, schartmann08}. Most models assume
that the dust is distributed in a `torus', i.e, a roughly axisymmetric structure with an finite inner radius, 
determined by the sublimation temperature of the grains. 
Smooth and clumpy torus geometries models with a range of parameters have been studied in the literature.

A consistent feature of most torus models is a broad peak in the continuum emission at MIR wavelengths
with a sharp dropoff beyond a certain break wavelength. The
MIR SED shape shows considerable variation in different models, due to silicate emission and
absorption features at 10 \mics\ which depend on the geometry of the dusty torus and its optical
depth in the MIR. The typical FIR SED is smoother and more uniform among 
most models. However, a small subset of models, characterised by
high MIR optical depths (or dust columns) and large torus sizes, peak at fairly
long wavelengths, approaching 60 \mics\ \citep{fritz06}. In such models, the intrinsic 
IR AGN SED is quite `cold' and a substantial fraction of the reprocessed 
AGN power can come out in the FIR.

In addition to torus emission, AGN light can also be reprocessed by dust on kpc scales, in the 
Narrow Line Region (NLR). 
NLR dust is typically cooler than torus dust and has a longer tail towards FIR wavelengths \citep{schweitzer08, mor09}.
The relative strength of this component with respect to the torus emission is a function of the covering factor of
NLR clouds, which tends to be quite low \citep[$\sim 0.1$;][]{mor09}, indicating that the NLR contribution to
the total AGN IR luminosity is likely to be small.

\begin{figure*}[t]
\includegraphics[width=\textwidth]{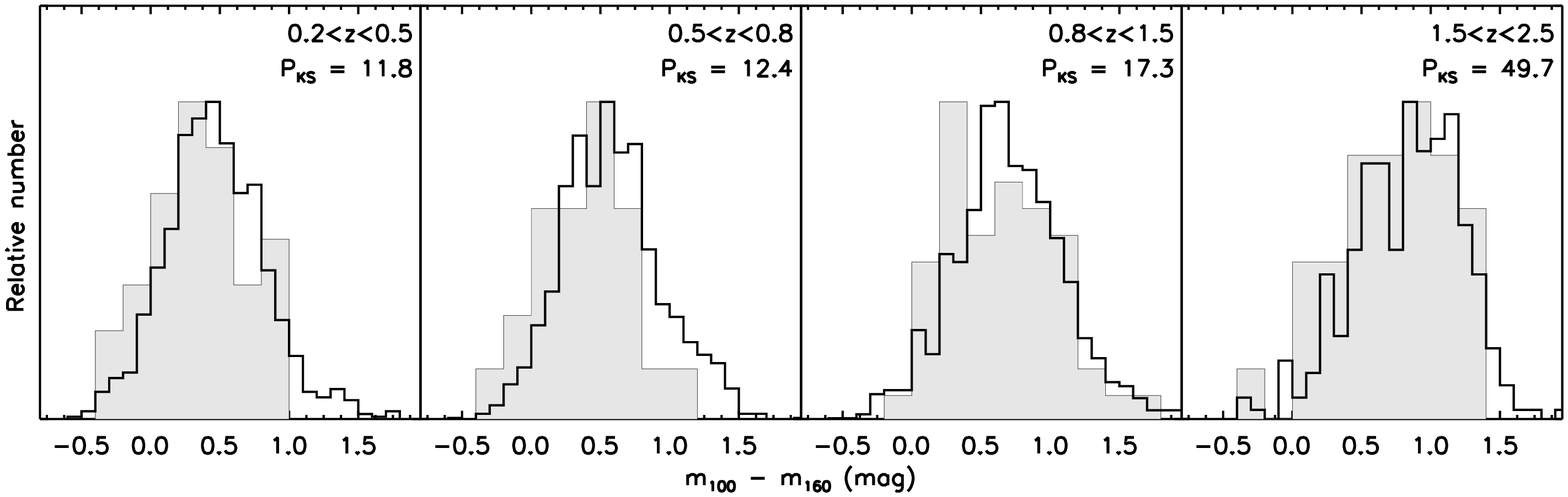}
\caption{Distributions of \mfir\ for COSMOS PACS-detected AGNs (shaded histogram) compared
to the general population of PACS detected galaxies in the COSMOS field. Each panel is a different redshift bin, indicated in the
upper right corner of the panel. Also in the upper right is a measure of the KS probability $P_{\rm{KS}}$
that the two distributions share a common parent distribution. The distributions are most different at $z<1$
but become more similar at higher redshift. Note, systems with likely AGN contamination are not 
excluded from this plot. 
}
\label{cosmos_colordist_all}
\end{figure*}

The theoretical approach of modelers is useful to understand the \emph{possible} range of intrinsic AGN SED shapes.
However, the `torus' and extended circumnuclear  environment of the SMBH, which governs the intrinsic SED,
has only been studied in a few nearby AGNs, due to the limited spatial resolution of MIR
and FIR imagers, and little is known of true variations in their geometry
and phase structure. 

Other works have taken an empirical approach towards calibrating 
intrinsic AGN SEDs. Such analyses rely on some knowledge of the range of
SEDs of star-forming galaxies, which are better constrained from studies of inactive systems. 
By subtracting IR templates of star-forming systems from SEDs of active galaxies, one may
reveal the residual excess which must arise from AGN-heated dust. 
Adopting this approach, \cite{netzer07}, in a study of PG QSOs with bolometric luminosities \lagn\
in the range of $10^{44}$-$10^{ 46.5}$ \ergs, derive a mean intrinsic AGN SED that shows
strong Si emission features and drops rapidly beyond $\sim 20$ \mics. The shape of the mean SED
does not appear to depend on the level of SF, but does shows
considerable intrinsic variation from object to object, especially at NIR-MIR wavelengths.

This work was extended recently by \cite{mullaney11a}, which explores intrinsic AGN SEDs in
a sample of 11 local low-to-moderate luminosity hard X-ray selected AGNs with low levels of SF.
Using a set of 5 templates which encompass most of the range
of star-forming galaxies, they isolate and model the AGN emission using a smooth broken power-law
model, with a sharp long-wavelength dropoff modeled as a modified black-body. The
empirical AGN SEDs from \cite{mullaney11a} show significant intrinsic variation both in MIR
and FIR, with some peaking as far red as 50 \mics. The study also uncovers a dependence
of SED shape on AGN luminosity, such that more luminous AGNs have, on average, a steeper
long-wavelength drop-off than lower luminosity AGNs. The long-wavelength slope of the
mean AGN SED from \cite{mullaney11a} is shallower than that of the mean QSO SED from
\cite{netzer07}, but this could be due to the lower characteristic AGN luminosity of the 
\cite{mullaney11a} sample. Having said this, detailed SED modeling of a sample
of Type I AGNs, covering a wide range in AGN luminosity and 
including contributions from NLR dust emission and hot dust from the outer Broad-Line Region, 
seems to suggest that the steep FIR dropoff in the \cite{netzer07} SED applies even to lower
luminosity systems and may be more universal \citep{mor11}. Given the recent nature of the latter paper,
we choose in this paper, to use the SEDs of \cite{mullaney11a} and \cite{netzer07}, but
caution the reader that this suite of SEDs may overestimate the contribution of the AGN
at FIR wavelengths.

\cite{mullaney11a} also compared their empirical AGN SEDs with those predicted from torus models. 
The empirical SEDs shapes show a much smaller scatter than those of the smooth torus models of \cite{fritz06}, 
but are more comparably matched to the expected range of the clumpy torus models of \cite{schartmann08}. 
This may arise simply from differences in the range of parameter space explored by the modelers --
the latter clumpy models are dependent on expensive 3-D modeling and explore a smaller
range of torus sizes and optical depths, while the \cite{fritz06} models with the largest variation are those
with large sizes and MIR optical depths.

\subsection{FIR colors of AGNs and inactive galaxies}

A simple approach towards exploring the importance of AGN contamination at FIR wavelengths is to compare
the FIR colors of AGNs against the general population of galaxies. The significantly warmer temperatures
of AGN-heated dust gives flat or falling SEDs at long IR wavelengths, which leads to bluer FIR
colors among galaxies with a substantial AGN component in the FIR. 

In the top panels of Fig.~\ref{cosmos_ccdiag}, we compare IR colors of PACS-detected X-ray AGNs 
and all PACS-detected galaxies in the COSMOS field in four redshift bins.
No explicit selection has been applied to the PACS-detected galaxies in this plot, except that
they have a well-defined spectroscopic or photometric redshift. 
The AGN points are colored by their hard-band X-ray luminosity (log \lhard) using a rainbow
color stretch from black (log \lhard$=42.0$ \ergs) to red (log \lhard$=45.5$ \ergs), more than 3.5 orders
of magnitude in AGN accretion luminosity. The low end of the \lhard\ range is a luminosity threshold 
which is popularly used to identify pure AGN samples, while the upper end overlaps the luminosity
range of moderately bright QSOs. 

The X-axis of Fig.~\ref{cosmos_ccdiag} is the 24 \mics\ to 160 \mics\ color ($m_{24\textrm{-}160}$), 
a MIR to FIR flux ratio, sensitive to warm AGN dust emission. 
At $z\gtrsim1$, the MIPS 24 \mics\ band is also influenced by silicate absorption, 
while, at $z\gtrsim2$, PAH emission contributes as well.
The FIR 100 \mics\ to 160 \mics\ color on the Y-axis (\mfir) measures the FIR continuum slope, down to the
rest-frame 30 \mics\ at $z=2.5$. 
The flux measurements in these two PACS bands are used to derive the rest-frame 60 \mics\ luminosity
in all four redshift bins. Therefore, understanding the level of contamination from AGN light
in these bands is particularly relevant to the later discussion on FIR trends with AGN luminosity.
Both colors are in units of magnitudes; therefore, smaller values (bluer colors) imply
systematically warmer SEDs. We will use the term `warmer' to refer to bluer IR colors to avoid
any confusion with the colors used in the plotting symbols.


Since the PACS bands probe different rest-frame wavelengths with redshift, we break 
the sample into our four standard redshift bins to reduce the effect of k-corrections.

From the Figure, one may notice that the AGNs
on average show warmer \mmir\
than normal (i.e, X-ray undetected) galaxies with FIR detections,
though the bulk of AGNs at any given redshift have colors that are 
consistent with the field population (see also the discussion of mid-IR spectra in section 8 of \cite{nordon12}).
The most luminous AGNs typically have the warmest 24-100 colors. 

The \mfir\ distributions of the AGNs and normal galaxies are more similar 
than their \mmir\ distributions \citep[also see][]{hatzimina10}, but 
some small differences are noticeable. In Fig.~\ref{cosmos_colordist_all}
we compare the distributions of \mfir\ between AGNs and normal galaxies.
At face value, we expect that the largest differences between the two sets of histograms should be at high
redshift. This is because the PACS 100 and 160 \mics\ bands 
trace increasingly shorter rest-frame wavelengths, making the AGN emission more prominent with redshift. 
However, Fig.~\ref{cosmos_colordist_all} shows that this is not the case. Quite surprisingly, the 
biggest difference between the distributions of AGNs and normal galaxies arises at low redshifts, especially
in the left two panels ($z<0.8$). 
The AGNs at these redshifts show a displacement to warmer FIR colors compared to 
the general galaxy population, with a warmer median value of the \mfir, 
and lack the long tail to cooler FIR colors that is seen among normal galaxies. 
In the two high redshift bins, the median values of AGNs and normal galaxies are more similar, though
the AGNs still show a stronger wing on the warm side of the distribution. 

To quantitatively assess the level of AGN contamination in the FIR,
we generate a suite of hybrid SEDs which combine 
AGN and SF templates picked to span the range of typical SED shapes found in real galaxies.
We adopt three AGN SEDs from \cite{mullaney11a} -- a mean AGN SED, a typical high luminosity
and a typical low luminosity SED \footnote{available from http://sites.google.com/site/decompir/} --
as well as the mean QSO SED from \cite{netzer07}, suitable for AGNs with QSO-like luminosities.
For the SF templates, we consider the five basis SEDs from \cite{mullaney11a} and add an M82-like
and Arp220-like SED from the SWIRE template library \citep{polletta07} to incorporate strongly star-bursting
and ULIRG-like systems. We create AGN-dominated 
hybrid SEDs by adding an AGN and a SF template scaled to
give a fractional contribution of AGN light $> 50$\%
at a rest-frame wavelength of 60 \mics. 

In each panel of Fig.~\ref{cosmos_ccdiag} we plot a shaded polygon that encompasses the
range of observed-frame colors of these hybrid templates after they have been 
appropriately redshifted to the upper and lower redshift bounds of each bin. The templates that
are fully AGN dominated are typically very warm in both \mmir\ and \mfir,
and many lie outside the plots. 
The cooler end of the polygons are populated by hybrids with $\sim 50\%$ AGN fractions. 

In all the panels, the polygons lie well away from the small black points which mark the location of the 
bulk of PACS detected galaxies. This is expected since most PACS detected galaxies are not likely to host
an active nucleus, even those that may have obscured AGNs missed by X-ray surveys. However, 
it is also evident that only a few  AGNs lie within these polygons at any redshift. By and large, most 
AGNs can safely be assumed to be
dominated by cool dust emission in the FIR. 

The handful of AGNs that do lie in the AGN-dominated polygons are typically quite luminous. 
In addition, there is a weak but general trend for the more luminous AGNs
to lie closer to the polygons in all the redshift bins. This implies that some small level of AGN contamination
does enter into the FIR, even around and beyond 60 \mics, and simply assuming that the rest-frame
60 \mics\ luminosity in AGNs is always attributable to star-formation is probably incorrect. It is clear,
however, that the vast majority of PACS detected AGNs are dominated by SF-related emission in the FIR.

Can a low level of AGN contamination in the FIR lead to the different \mfir\ distributions between
AGNs and inactive galaxies? Two features of Fig.~\ref{cosmos_ccdiag} suggest otherwise.
Firstly, note that the shaded area
of strong AGN contamination is most offset from the the locus of the data points in the lowest
redshift panel and closest to the locus in the highest redshift panel. This is simply because
the effects of AGN contamination on \mfir\ are least pronounced at low redshifts, for reasons already
discussed. Yet, it is in these low redshift bins where the most significant differences in color distributions
are seen in Fig.~\ref{cosmos_colordist_all}. While the exclusion of AGN dominated
objects does make the distributions more similar, 
substantial differences still remain, especially the lack of a cool wing to the AGN \mfir\ distribution.
Secondly, close examination of Fig.~\ref{cosmos_ccdiag} reveals that objects that 
make up the spread to warmer FIR colors are not always luminous AGNs, as one would expect
if contamination was the reason for the spread. 
In addition, the part of the color-color diagram occupied by `warm' AGNs at low redshifts
also contains some X-ray undetected objects, implying that AGN activity is not a necessary prerequisite
for having warm FIR colors.
The lack of a close relationship between FIR color and AGN luminosity,
coupled with an inverted trend with redshift, suggests that the offsets are not related to 
AGN activity, but are intrinsic to the population of galaxies that host AGNs. 

In \cite{santini11}, we show that X-ray AGNs at $z<1$ exhibit slightly enhanced mean SFRs
compared to normal galaxies of the same stellar mass, while the offset declines or disappears
at higher redshifts. It is possible that the very processes that are responsible for the enhancement
of mean SF activity in lower redshift AGNs may also affect their typical FIR colors.  We discuss
this further in \S5.

We turn now to AGNs that were not detected in both PACS bands. They
account for the vast majority of X-ray selected AGNs ($\approx 70\%$ in the GOODS fields, more than 
90\%\ in the COSMOS field). PACS-undetected AGNs are less star-forming, 
and therefore, on average, more AGN dominated at a given AGN luminosity
than the PACS-detected AGNs. We can study the FIR properties of PACS-undetected AGNs only
through their mean stacked signal. 
In the lower panels of Fig.~\ref{cosmos_ccdiag}, we compare the mean colors
of the PACS undetected AGNs to the PACS detected AGNs. 
The stacked points come from individual measurements in all three fields, while the black points
here are only from the COSMOS dataset and serve only to place the stacked points in context.
The stacks were restricted to objects that were detected in the 24 \mics\ band, in order to 
allow an accurate estimate of the mean 24 \mics\ flux. 
As before, we include the polygons that delineate regions in the diagrams occupied by AGN-dominated
SEDs at 60 \mics\ rest. 

By and large, the colors of the PACS stacked AGNs lie outside the shaded polygons, implying that
the stacked signal is not dominated at 60 \mics\ by AGN light. However, as expected
they tend to lie closer to the polygons than the bulk of detected AGNs, clustering around
the right edges where templates with AGN fractions of
a few tens of percent would lie. Some or most of the stacks in any given redshift bin lie within the 
region of AGN dominance. In general, the scatter of stacked points lies closer to the polygons
in the higher redshift bins, as expected if the shift to warmer colors is a result of AGN contamination.

In addition, the stacks of the more luminous AGNs (redder points) have consistently warmer
colors. Most of the variation with AGN luminosity happens
through the \mmir\ color. Systematic changes in \mfir\ are not clearly discernable and the typical
FIR colors of the stacked AGNs are consistent with the typical colors of the detected AGNs.
This highlights the fact that FIR colors are not greatly affected by increases in AGN luminosity,
unlike the accretion sensitive MIR SED. We conclude that, at lower redshifts ($z<1$)
AGN contamination in the FIR is minor in both PACS detected and stacked samples and very 
rarely gets above 50\%. However, at higher redshifts out to $z=2.5$, the AGN can substantially
affect \mfir\ and, therefore, our estimate of the rest-frame 60 \mics\ luminosity.

\begin{figure*}[ht]
\includegraphics[angle=90,width=\textwidth]{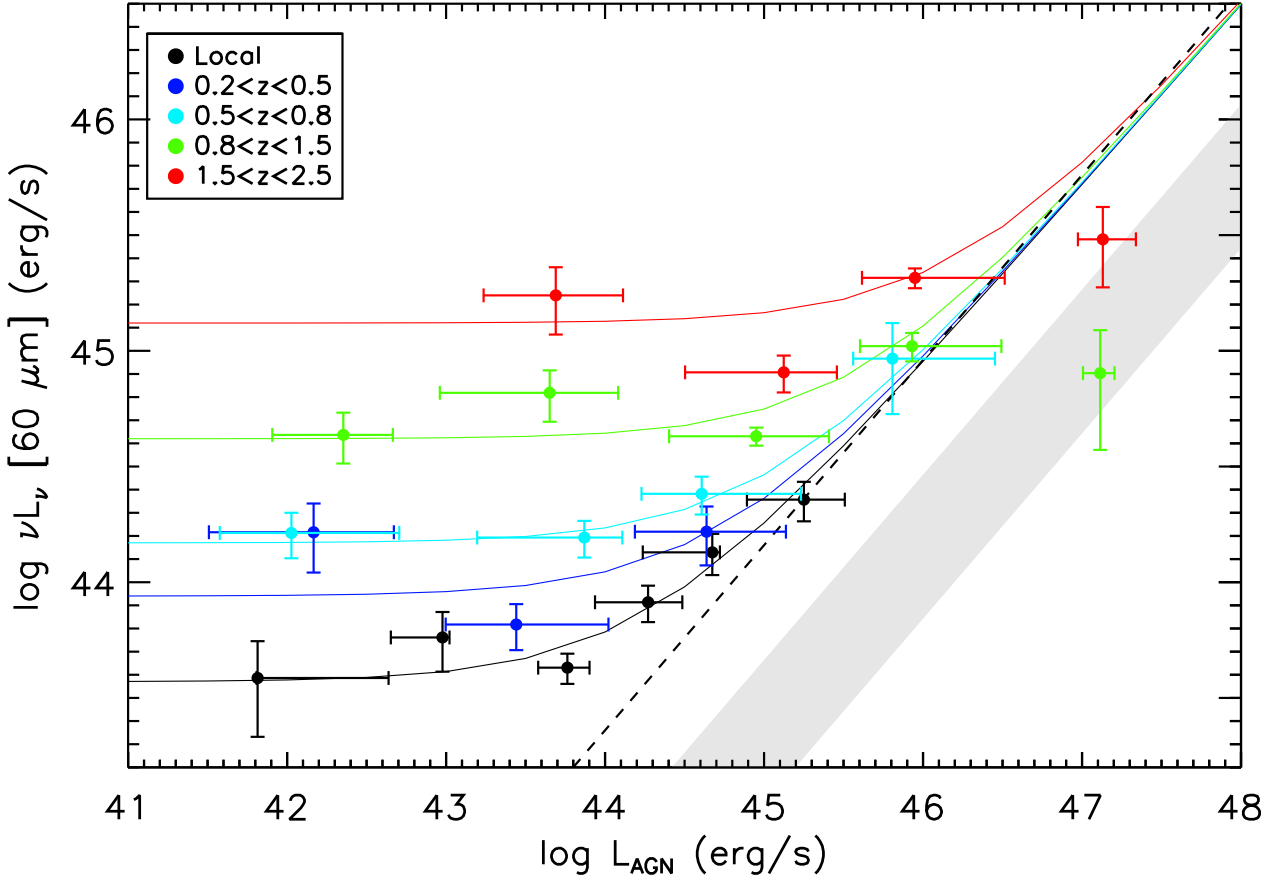}
\caption{Mean \lums\ (\lfir) vs. $L_{AGN}$ of X-ray selected AGNs in 5 different redshift 
bins from the local Universe to $z=2.5$. The colored data points are combinations of mean
measurements in 3 PEP fields: GOODS-N/S and COSMOS, while the black data points
come from our analysis of the SWIFT BAT sample. The solid colored lines
are functional fits to the mean measurements, as described in \S4.1.1. The dashed line is the
correlation line shown by AGN-dominated systems in \cite{netzer09}. The shaded region
corresponds to the approximate 1$\sigma$ range exhibited by empirical pure-AGN SEDs.
At low redshifts, a strong change in the mean trend exists as a function of \lagn, which 
disappears at high redshifts. The mean \lfir\ of low-luminosity X-ray AGNs increases 
monotonically with redshift, mirroring the increase in the mean SFR of massive galaxies 
across redshift.
}
\label{shao_plot_main}
\end{figure*}

\section{Results}

Using measurements of \lfir, the mean rest-frame 60 \mics\ luminosity from all 3 fields combined (Appendix B) , 
we will now study the relationship between FIR emission and SMBH accretion to $z=2.5$. We first look at mean
trends that come from combining fluxes from detections and stacks in the PACS bands to get a single
mean value for \lfir\ in each bin in redshift and X-ray luminosity. In the following sections,
we extrapolate \lhard\ to a bolometric AGN luminosity (\lagn) using the following relation, which is based
on Eqn.~5 from \cite{maiolino07} and taking a ratio of 7 between the luminosity at 5100\AA\ and \lagn\ \citep{netzer07b}.

\begin{equation}
\log L_{AGN}  =  \frac{(\log L_{X} - 11.78)}{0.721}  + 0.845
\end{equation}

\noindent where \lhard\ (the 2-10 keV band X-ray luminosity) and \lagn\ are in \ergs.

\begin{table*}
\caption{Mean Rest-frame 60 $\mu$m luminosities in bins of Redshift and Hard-band X-ray Luminosity.}              
\label{table1}      
\centering                                      
\begin{tabular}{c c c c c c}          
\hline\hline                        
Redshift bins & \multicolumn{4}{c}{Bins in De-absorbed 2-10 keV band X-ray luminosity \lhard} \\    
\hline                                   
COSMOS &  41 -- 42 & 42 -- 43 & 43 -- 44 & 44 -- 45 & 45 -- 45.5 \\
\hline  
0.2 -- 0.5 &  43.17 -- 44.04   (18) &  43.72 -- 43.93  (100) &  44.07 -- 44.33  (34) &   ---  (2)   &   ---  (0)  \\
0.5 -- 0.8 &   ---  (1)    &  44.03 -- 44.24  (73)  &  44.19 -- 44.42  (123) &  44.73 -- 45.12  (13)     &   ---  (0)  \\
0.8 -- 1.5 &   ---  (0)    &  44.39 -- 44.96  (28)    &  44.59 -- 44.68   (434)  &  44.97 -- 45.11  (134) &  44.57 -- 45.09  (4)  \\
1.5 -- 2.5 &   ---  (0)    &   ---  (0)    & 44.78 -- 44.98   (152)   & 45.30 -- 45.39   (295) &  45.27 -- 45.62  (23)  \\
\hline                                 
GOODS-S &  41 -- 42 & 42 -- 43 & 43 -- 44 & 44 -- 45 & 45 -- 45.5 \\
\hline                                   
0.2 -- 0.5 &  44.02 -- 44.49  (11) &   ---  (1) &   ---  (1) &   ---  (0) &   --- (0)   \\
0.5 -- 0.8 &  43.90 -- 44.14  (30) &  44.04 -- 44.45  (19) &  43.94 -- 44.35  (11) &   ---  (1) &   --- (0)   \\
0.8 -- 1.5 &  43.99 -- 44.61 (18) &  44.40 -- 44.93  (27) &  44.43 -- 44.75  (23) &  44.88 -- 45.15  (3) &   ---  (0)  \\
1.5 -- 2.5 &   ---  (3) &  45.02 -- 45.39  (20) &  44.66 -- 45.05  (32) &  44.87 -- 45.13  (5) &   ---  (0)  \\
\hline                                 
GOODS-N &  41 -- 42 & 42 -- 43 & 43 -- 44 & 44 -- 45 & 45 -- 45.5 \\
\hline                                   
0.2 -- 0.5 &  43.88 -- 44.41  (12) &  42.95 -- 43.81  (5) &   ---  (2)  &   --- (0)  &   ---  (0)  \\
0.5 -- 0.8 &  44.23 -- 44.59  (22) &  44.03 -- 44.34  (10) &  44.36 -- 44.67  (5) &   --- (0)  &   --- (0)   \\
0.8 -- 1.5 &  44.53 -- 44.77  (33) &  44.69 -- 44.98  (47) &  44.51 -- 44.71  (39) &  44.28 -- 44.75  (6) &   ---  (0) \\
1.5 -- 2.5 &   ---  (3) &  44.89 -- 45.42  (16) & 44.72 -- 45.11  (44) &  44.08 -- 45.27  (5) &   ---  (0)  \\
\hline                                          
\end{tabular}
\tablefoot{ 
Mean Rest-frame 60 $\mu$m luminosities are in units of log \ergs.  
Numbers in parentheses indicate the number of galaxies that contribute to the mean measurement in each redshift
and X-ray luminosity bin.
}
\end{table*}

\subsection{Mean trends: L$_{60}$ against L$_{AGN}$} 

In Fig.~\ref{shao_plot_main}, \lfir\ is plotted against \lagn. The plotted
\lagn\ is the median value for objects in all three fields in AGN luminosity bin, while
error bars in \lagn\ show the interpercentile range in AGN luminosity containing 80\% of the sample -- 
typical errors in AGN luminosity are smaller than the error bars shown here. Different colors are used to represent
measurements in the different redshift intervals. All the mean measurements from the Herschel fields 
have been tabulated in Table \ref{table1}. In the plot, we include as well data for the local sample of Swift BAT AGNs 
(black points and line). Included as well in the figure is the relation from \cite{netzer09} (dashed line)
and colored solid curves which come from our fits in \S4.1.1. This plot may be compared directly to Fig.~6 of \cite{shao10}
and is analogous to Fig.~5 in \cite{mullaney11b}. 

We also include a shaded zone which shows the range in the
plot occupied by pure AGNs (no star-formation) with the SEDs described in \S3.1. These
SEDs are expected to span the majority of true AGN SED shapes in nature.
The tight correlations between \lhard\ and 12.3 \mics\ IR luminosity among 
local AGNs \citep{gandhi09} are used to link the bolometric AGN luminosity (derived
from \lhard) and the FIR luminosity (through the AGN IR SEDs). While, at low AGN luminosities, torus
emission will not contribute measurably to the mean FIR luminosity, it plays an increasingly
greater role at high AGN luminosities. 
Indeed, quite surprisingly, our \lfir\ measurements for the most luminous
AGNs at $z>1$ are consistent with their mean SEDs having low or no star-formation.
On the other hand, the average of AGN+host SEDs should be above the pure AGN SED if even some objects 
have star formation. This may indicate that our adopted intrinsic AGN SEDs are too FIR-bright, as suggested
by \cite{mor11}.

At redshifts below $z=0.8$, the variation of \lfir\ with \lagn\ shows a consistent behavior. At low AGN luminosities,
\lfir\ is weakly correlated or uncorrelated with AGN luminosity, and approximately 
constant over two orders of magnitude in \lagn. This constant FIR luminosity,
which we call \lfirm, is redshift dependent and rises by a factor of $\approx 3$ between 
$z=0$ and $z=0.65$ (the mean redshift of galaxies in the $0.5<z<0.8$ bin). 

At higher AGN luminosities, a change is seen. Beyond a certain transition in \lagn, \lfir\ begins to correlate strongly with \lagn.
This is most evident in the local BAT sample and among AGNs in the $0.5<z<0.8$ bins,
which both span a sufficiently large range in \lagn\ to bring out this trend. The slope of the correlation is
comparable in both redshift bins. 
The transition AGN luminosity, which we call \lagnc, increases with redshift, 
by about an order of magnitude between $z=0$ and $z=0.8$. 

While lower redshift AGNs show a particular relationship between \lfir\ and \lagn, the situation 
appears to change at higher redshifts ($z>0.8$). The correlation between \lfir\ and \lagn\ is weak or absent. 
The overall trend is much flatter than for AGNs at lower redshifts, implying that beyond $z\sim1$ the FIR 
emission of the AGN hosts is quite independent of the accretion luminosity of the nucleus. \lagnc\ is not well 
characterised at these redshifts.

Having said this, we note that these inferences come primarily from measurements 
of the highest luminosity AGNs in our sample, which are significantly lower than expected from 
an extrapolation of the low redshift correlation. 
This measurement in the $0.8<z<1.5$ redshift bin is rather uncertain since it depends on 
a combination of only four objects and is subject to large statistical uncertainties. 
The measurement in the $1.5<z<2.5$ bin is more robust. We have individually examined the optical and
IR images of these luminous AGNs, as well as the X-ray spectral fits, to ensure that they 
are not affected by incorrect multi-wavelength associations or large errors. 
Therefore, we believe that the flattening of the \lfir\--\lagn\ relationship among luminous AGNs
at high redshifts is a real effect. 

The increase in the mean \lfir\ of low luminosity AGNs continues to $z=2.5$. The change
in \lfirm\ accelerates between $z\sim1$ and $z\sim2$, to reach \lfirm$\sim10^{45}$ \ergs\ at
$1.5<z<2.5$. This is somewhat at odds with the increase in the star-formation density of field galaxies
with redshift, which typically rises sharply to $z\approx1$ and flattens out towards higher redshift \citep{hopkins06}. 
However, AGN are typically hosted by massive galaxies \citep{brusa09, xue10, rosario12}
which have star-formation histories that are different from
the bulk of star-forming galaxies at these redshifts \citep{perezgonzalez08}. Star-formation `downsizing'
requires that such massive galaxies go through most of their star-formation 
around or earlier than $z\sim2$ and contribute less to the integrated FIR
luminosity density of the Universe at $z<1$ than at higher redshifts. Therefore, we would expect
AGN hosts to have a higher rate of increase of \lfirm\ at $1<z<2$ than at $z<1$ \citep{santini11}.

\subsubsection{Fits to the \lfir\ - \lagn\ trend}

\begin{table*}
\caption{Functional fits to the Trends between \lfir\ and \lagn\ in bins of Redshift}              
\label{table2}      
\centering                                      
\begin{tabular}{c c c c c c}          
\hline\hline                        
\multicolumn{6}{c}{Slope of Regression Line $\alpha$ left free} \\    
\hline
Redshift bins & \lfirm\ \tablefootmark{1}& $\alpha$ \tablefootmark{2} & \lagnc\ \tablefootmark{3} &$\overline \chi^{2}$ & $\overline \chi^{2}$ (single line fit) \tablefootmark{4}\\   
\hline
Local        & $43.56\pm0.12$  & $0.75\pm0.22$  & $44.23\pm0.13$  & 1.53   & N/A   \\
0.5 -- 0.8  & $44.18\pm0.08$  & $0.84\pm0.30$  & $44.94\pm0.33$  &  0.41  & N/A   \\
0.8 -- 1.5  & $44.60\pm0.09$  & $0.58\pm0.18$  & $45.99\pm0.64$  &  7.58  &  7.65  \\
1.5 -- 2.5  & $45.08\pm0.08$  & $0.53\pm0.25$  & $46.46\pm0.82$  &  7.57  &  6.97  \\
\hline                                          
\multicolumn{6}{c}{Slope of Regression Line $\alpha$ fixed at 0.78} \\    
\hline
Redshift bins & \lfirm\ & $\alpha$ & \lagnc\ & $\overline \chi^{2}$ & \\   
\hline
Local        & $43.57\pm0.08$  & $0.78$  & $44.27\pm0.11$  & 1.14   & ---   \\
0.2 -- 0.5  & $43.94\pm0.09$  & $0.78$  & $44.82\pm0.41$  &  6.47  & ---   \\
0.5 -- 0.8  & $44.17\pm0.07$  & $0.78$  & $44.89\pm0.22$  &  2.42  & ---   \\
0.8 -- 1.5  & $44.62\pm0.04$  & $0.78$  & $45.98\pm0.17$  &  5.61  &  ---  \\
1.5 -- 2.5  & $45.12\pm0.06$  & $0.78$  & $46.51\pm0.21$  &  5.39  &  ---  \\
\hline    
\end{tabular}
\tablefoot{ 
See \S4.1.1. 
\tablefoottext{1}{Mean \lfir\ of low-luminosity AGNs}
\tablefoottext{2}{Slope of Regression line valid for luminous AGNs}
\tablefoottext{3}{Transition AGN luminosity: \lfir\ is significantly correlated with \lagn\ at $z<1$ beyond this AGN luminosity}
\tablefoottext{4}{A single straight line fit between log \lagn\ and log \lfir with a variable slope and normalisation.}
}
\end{table*}

We model the trend between log \lfir\ and log \lagn\ in each of the five redshift bins as a combination of a
flat line (a constant value \lfirm) and a straight line with a positive non-zero slope $\alpha$. 
The constant value is determined by the mean SFR of low-luminosity AGN hosts and is unrelated to accretion activity.
We capture the correlation between log \lagn\ and log \lfir\ as a linear relationship, essentially a regression line.
In total, three parameters define the fit: \lfirm, $\alpha$ and the intercept of the regression line. The intercept also
defines \lagnc, which we take to be the value of \lagn\ where the regression line ordinate = \lfirm, i.e,
the AGN luminosity where SF activity that is correlated with BH growth 
starts to dominate the total SFR of the host galaxy.

We fit the measured data using a non-linear least-squares fitting procedure (CURVEFIT in IDL). 
Results are tabulated in Table \ref{table2}. In the two lower redshift bins where a good fit is possible (local and $0.5<z<0.8$)
we arrive at consistent values for the slope of the regression line ($\alpha \approx 0.75\pm0.23$). \lfirm\ increases 
strongly with redshift, as expected. 

Among the two higher redshift bins, the transition is not well-defined. A best-fit slope of the regression line
is lower ($\alpha \sim 0.5\pm0.23$) and \lagnc\ is about 1.5 dex higher than at lower redshifts. However, the
quality of the fit is rather poor (reduced $\chi^{2} \gtrsim 3.5$) in both these redshift bins and, in fact, a single
straight line with a variable slope fits the data points just as well.

A correlation between AGN luminosity (or mass accretion rate) and host galaxy SF rate has been reported
by several studies, for local low-luminosity Seyferts \citep{netzer09, diamondstanic11aph} and high
redshift AGNs and QSOs \citep{lutz08, hatzimina10, bonfield11}. \cite{netzer09} compared the SFRs of
local SDSS Type II and LINER AGNs and find that they lie on the same relationship as luminous PG QSOs
\citep{netzer07} and the mm-bright QSOs at $z\sim2$ \citep{lutz08}. The slope of this relationship is approximately
0.8, quite consistent with our own estimate of $\alpha$. The relationship in \cite{netzer09} extends linearly
to low values of AGN luminosity (\lagn$<10^{43}$ \ergs), while we find a very definite turnover to a flat relationship
below \lagnc, most likely since \cite{netzer09} specifically concentrated on AGN-dominated galaxies.

If we assume that the slope
of the regression line is constant across redshift, as suggested by \cite{netzer09}, 
then the fits to our lower redshift data points gives the following relation:

\begin{equation}
\log \frac{L _{60}}{10^{44} \textrm{erg s}^{-1}} = \log  \left( \frac{L _{AGN}}{6.3\times10^{44} \textrm{erg s}^{-1}}\right)^{0.78}
\end{equation}

We rerun our fits now fixing the regression line to Eqn.~2 and derive a second, more refined set of values for \lfirm, which
are also listed in Table \ref{table2}. In Fig.~\ref{shao_plot_main}, we include lines which show the \emph{expected} 
trends in each of our five redshift intervals as solid colored lines. 
The data points for AGNs in the three lower
redshift bins  follow the trends quite well, since the trends were calibrated
on these points. At high redshifts, the high \lagn\ measurements lie systematically off the expected trend,
by as much as 3-4$\sigma$ for AGNs
at $z\sim1$ and by around $2\sigma$ at $z>1.5$. \lfir\ for these highest luminosity AGNs appear to be
comparable to, or perhaps slightly higher than, \lfirm\ at their corresponding redshifts. In other words, the mean 
FIR luminosity of the brightest AGNs at $z\gtrsim1$ are not very different from those of low-luminosity AGNs
or normal massive galaxies at these redshifts. This may be contrasted with the rather drastic change in the
\lfir\ with \lagn\ among AGNs at lower redshifts. This suggests that there is a change in the relationship
between SF and AGN activity between luminous AGNs at low and high redshifts, which becomes important
around $z\sim1$.

\begin{figure*}[t]
\includegraphics[width=\textwidth]{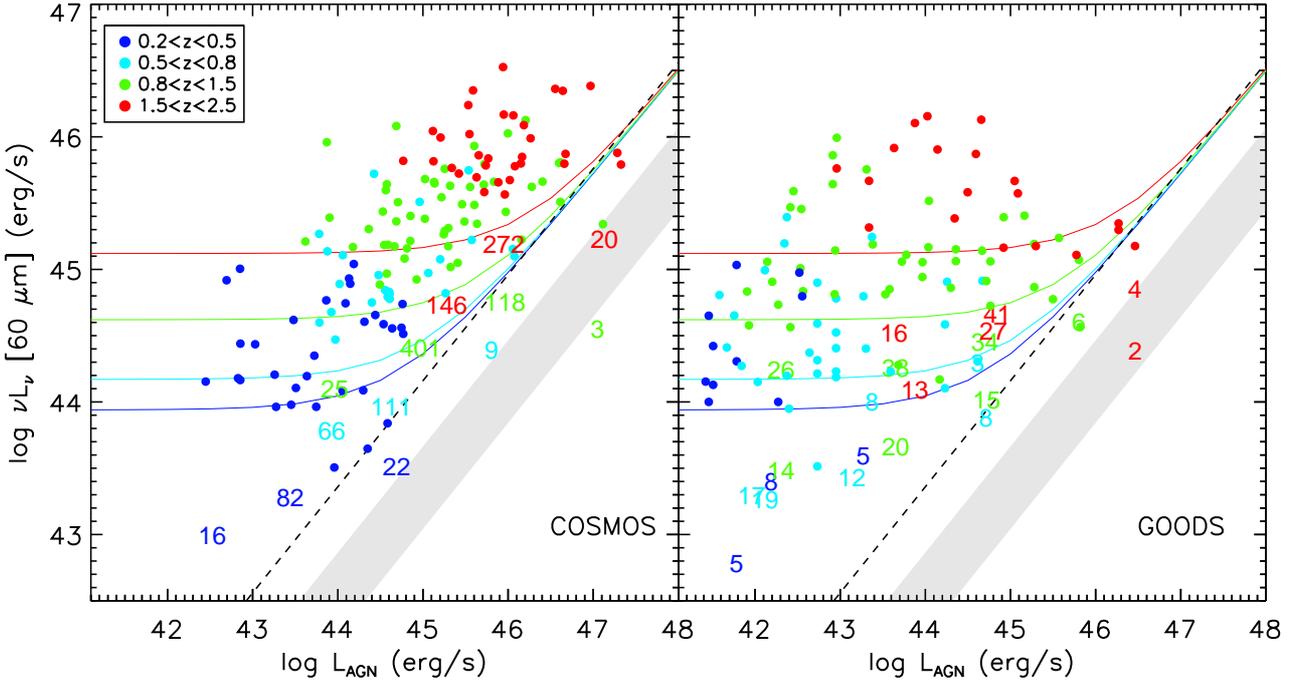}
\caption{\lums\ vs. $L_{AGN}$ of X-ray selected AGNs in 4 redshift
bins from $z=0.2$ to $z=2.5$. Galaxies in the GOODS and COSMOS fields have been
separated into right and left panels respectively, as they differ greatly in both X-ray and
PACS depths. Colored points are AGNs detected in both bands. Colored numbers 
show the location of stacked measurements and the number of objects that go into
the stack. Dashed and solid lines, as well as shaded regions, are the same as in Fig~\ref{shao_plot_main}.
Note the apparent correlation in the colored points in the left panel, which
is absent in the right panel. The correlation is primarily driven by selection effects
inherent with wide and shallow surveys.
}
\label{det_vs_stack}
\end{figure*}

\subsection{Distributions of L$_{60}$ against L$_{AGN}$} 

Till now, we have concentrated on the mean FIR properties of the AGN population. Underlying these
mean measurements is a distribution of FIR luminosities which can have considerable intrinsic scatter.
We turn now to an examination of the \lfir\ distribution by considering separately properties of
AGNs both detected and undetected in the PACS imaging.

In Fig.~\ref{det_vs_stack}, we plot \lums\ against \lagn\ for PACS-detected AGNs on a diagram similar
to that used for the mean measurements in Fig.~\ref{shao_plot_main}. We also
indicate the location in this diagram of the stacked measurements for PACS undetected AGNs
using colored numbers which list the number of objects that go into a given stack. Since the X-ray
and PACS depths of the COSMOS field is considerably lower than the GOODS fields, we
make separate panels in the Figure for COSMOS and the combined GOODS fields. 

If we concentrate first on PACS detected AGNs, a clear correlation can be seen between
AGN luminosity and \lums\ in the COSMOS dataset if we consider AGNs at all redshifts together.
While this correlation has a slope that matches that of the regression line that we derive above
for high-luminosity AGNs, it is offset from the line by approximately an order of magnitude in \lums.
The correlation disappears, however, when one considers PACS detected AGNs in the GOODS
fields, even though there is a large degree of overlap in \lagn\ between the two samples. 
The correlation among COSMOS AGN is not real -- it is driven primarily by strong selection effects
in the AGN sample with redshift. At high redshifts, the rare X-ray {\it and} FIR luminous AGNs 
which are not found in low redshift bins due to smaller survey volumes start to be 
represented significantly in the COSMOS sample. This effect can be seen even within a single large
redshift bin (for e.g, the green points at $0.8<z<1.5$) though the apparent
correlation is weaker. The much smaller survey volume of the combined
GOODS fields ensures that luminous AGNs are under-represented at all redshifts
allowing for a more uniform AGN sample, which weakens the aforementioned
selection effects and, by extension, the apparent correlation between \lums\
and \lagn. We suggest that relationships between star-formation 
and nuclear activity in AGNs in the literature that are based on individual detections from shallow, large-area
surveys may be strongly affected by such selection effects \citep[e.g.,][]{hatzimina10}.

In general, the stacked measurements are up to an order of magnitude less luminous than the
weakest PACS-detected sources (see also Fig.~\ref{z_vs_l60}).
A substantial fraction of AGNs, especially at lower redshifts, are hosted
by quenched galaxies \citep{santini11}. Since star-formation is very weak to negligible in 
such hosts, they are expected to lie well below the PACS detection limit at all redshifts
and therefore contribute only to the stacked measurements. 

\begin{figure}[ht]
\includegraphics[angle=90,width=\columnwidth]{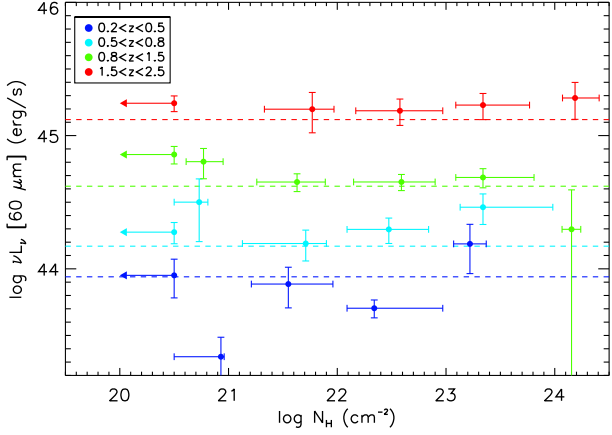}
\caption{\lfir\ vs. $N_H$ in 4 different redshift bins in the redshift range $0.2<z<2.5$. 
The dashed lines show \lfirm\ in each redshift bin, taken from 
Fig.~\ref{shao_plot_main}.  All objects with $N_H$ consistent with Galactic
absorption contribute to the point at $N_{H} \leq 10^{20.5}$ cm$^{-2}$, irrespective
of the different Galactic absorption towards the three different fields.
 \lfir\ is relatively flat across a large range in N$_{H}$ with no clearly 
 discernible trends, especially among well-measured points.
As discussed in \S4.3, covariances between \lhard\ and $N_H$, driven by
selection effects in the population of obscured AGNs, can
affect trends in this figure, but the general results shown here should
be relatively robust, since we combine measurements from fields of
very different X-ray depths.
}
\label{nh_trends}
\end{figure}

\subsection{Mean trends: $N_H$ against L$_{AGN}$} 

We turn now to the variation of \lfir\ with X-ray obscuration, parameterised by the column
density of absorbing hydrogen $N_H$. The major merger model for AGN fueling and evolution predicts
that most BH growth occurs when the AGN is enshrouded by a dense dusty obscuring
envelope at the center of a merger remnant \citep[e.g.,][]{sanders88}. If mergers are responsible for fueling most
of the X-ray AGN population, as some evolutionary models propose \citep[e.g.,][]{somerville08},
there should be definite trends between the obscuration of AGNs and their mean SFR.

We explore this in Fig.~\ref{nh_trends}, where we plot \lfir\ against $N_H$ in the four
redshift bins that we used in earlier figures. To ensure a clean AGN sample,
only sources with \lhard$> 10^{42}$ \ergs\ were used to make this plot.
At each redshift bin, we split the AGNs into six bins in obscuration:
 log $N_H$ (cm$^{-2}$) of $<20.5$, 20.5-21, 21-22, 22-23, 23-24, $>24$.
The plotted value of $N_H$ for each data point is the median value
for all AGNs in that bin and the error bars show the interpercentile range containing 80\% of the sample.
Measurements from all three fields have been combined in this Figure. 

In general, the trend between \lfir\ and $N_H$ is weak.  
At $0.5<z<0.8$,  \lfir\ appears to increase slowly with $N_H$ by about 0.4 dex across 3
orders of magnitude in the obscuring column, though unobscured AGNs also show a higher
mean \lfir. The measurement for the intermediate bin in $N_H$ is consistent
with the \lfirm\ of the AGN population at this redshift, while bins with higher obscuration 
are slightly enhanced compared to this value. This trend appears to be driven more
by selection effects in \lhard--$N_H$ space than by true correlations between \lfir\ and $N_H$
(see Fig.~\ref{lx_vs_nh} and associated discussion) and is not particularly significant .
At $0.8<z<1.5$ and $1.5<z<2.5$, the increase of \lfir\ with $N_H$ is weaker, 
consistent with a flat trend.

At every redshift, a subset of AGNs have $N_H < 10^{20.5}$ cm$^{-2}$, 
consistent with a very low or pure Galactic obscuration. In general, \lfir\ for these objects
is comparable to or higher than those of more obscured AGNs.  

\begin{table*}
\caption{Mean Rest-frame 60 $\mu$m luminosities in bins of Redshift and X-ray Obscuring Column.}              
\label{table4}      
\centering                                      
\begin{tabular}{c c c c c c c}          
\hline\hline                        
Redshift bins & \multicolumn{5}{c}{Bins in Column Density of Obscuring Hydrogen $N_H$ (log cm$^{-2}$)} \\    
\hline                                   
COSMOS &  19 -- 20.5 & 20.5 -- 21 & 21 -- 22 & 22 -- 23 & 23 -- 24 & 24 -- 25\\
\hline  
0.2 -- 0.5 & 43.78 -- 44.07 (45) & 43.12 -- 43.49 (4) & 43.71 -- 44.01 (36) & 43.90 -- 44.17 (46) & 43.96 -- 44.33 (4) &        ---        (0) \\
0.5 -- 0.8 & 44.20 -- 44.45 (77) & 44.20 -- 44.68 (5) & 44.06 -- 44.31 (38) & 44.02 -- 44.37 (76) & 44.61 -- 45.08 (13) &        ---        (0) \\
0.8 -- 1.5 & 44.74 -- 44.89 (248) & 44.86 -- 45.22 (13) & 44.58 -- 44.72 (88) & 44.61 -- 44.76 (212) & 44.71 -- 45.00 (37) &        ---        (2)  \\
1.5 -- 2.5 & 45.18 -- 45.31 (289) &        ---        (2) & 45.02 -- 45.33 (35) & 45.09 -- 45.31 (84) & 45.16 -- 45.41 (58) & 45.13 -- 45.49 (3)   \\
\hline                                 
GOODS-S &  19 -- 20.5 & 20.5 -- 21 & 21 -- 22 & 22 -- 23 & 23 -- 24 & 24 -- 25\\
\hline                                   
0.2 -- 0.5 &        ---        (0) &        ---        (0) &        ---        (0) &        ---        (0) &        ---        (1) &        ---        (0) \\
0.5 -- 0.8 & 43.68 -- 44.69 (7) &        ---        (2) & 43.59 -- 44.33 (3) & 43.87 -- 44.31 (11) & 43.80 -- 44.07 (7) &        ---        (1)   \\
0.8 -- 1.5 & 44.49 -- 45.33 (8) &        ---        (2) & 43.96 -- 44.47 (8) & 44.38 -- 44.83 (19) & 44.18 -- 44.72 (14) &        ---        (2) \\
1.5 -- 2.5 & 44.82 -- 45.40 (8) &        ---        (0) &  $<$44.59 (6) & 43.88 -- 44.78 (17) & 44.54 -- 45.23 (23) &  $<$44.60 (3)  \\
\hline                                 
GOODS-N &  19 -- 20.5 & 20.5 -- 21 & 21 -- 22 & 22 -- 23 & 23 -- 24 & 24 -- 25\\
\hline                                   
0.2 -- 0.5 &        ---        (2) &        ---        (0) &        ---        (1) & 43.39 -- 43.54 (3) &        ---        (0) &        ---        (1)  \\
0.5 -- 0.8 & 44.06 -- 44.29 (3) &        ---        (1) &        ---        (2) & 44.22 -- 44.51 (5) &        ---        (2) &        ---        (2) \\
0.8 -- 1.5 & 44.74 -- 45.12 (19) & 44.13 -- 44.44 (3) & 44.04 -- 45.29 (11) & 44.40 -- 44.64 (22) & 44.51 -- 44.68 (30) & 42.66 -- 44.59 (7)  \\
1.5 -- 2.5 &  $<$ 45.04 (11) &        ---        (1) &  $<$ 45.44 (4) & 44.73 -- 45.44 (14) & 44.82 -- 45.23 (27) & 44.86 -- 45.36 (8) \\
\hline                                          
\end{tabular}
\tablefoot{
Mean Rest-frame 60 $\mu$m luminosities are in units of log \ergs.  
Numbers in parentheses indicate the number of galaxies that contribute to the mean measurement in each redshift
and $N_H$ bin.
}
\end{table*}

We note here our AGN sample differs considerably from the luminous QSO samples for which a link between 
X-ray obscuration and SFR has been previously been suggested \citep{page01, page04, stevens05}. 
The bulk of our AGNs are typically below the knee of the X-ray LF, while those studies concentrated on 
special populations of X-ray absorbed optical broad-line QSOs. In addition, the typical AGN 
luminosity of our X-ray sample increases with redshift (Fig.~\ref{lx_vs_nh}). At $z<1$, most
AGNs that contribute to Fig.~\ref{nh_trends} are of low and moderate luminosity, while
at $z>1$, they are increasingly luminous AGNs where merger-driven processes may be
expected to be most relevant.

Another simple, model-independent way to 
broadly separate obscured and unobscured AGNs is through 
their Hardness Ratio ($HR$), defined as  $HR = (H-S)/(H+S)$     
where $H$ are the photon counts in the hard band (2-10 keV) and $S$ are the photon counts
in the soft band (0.5-2 keV). If the typical SFRs of obscured AGNs were significantly
enhanced, as suggested by models of AGN fueling by gas-rich major mergers,
then we expect them to be more frequently detected by PACS (or other FIR surveys)
compared to unobscured AGNs \citep[e.g.,][]{page11}.

\begin{figure}[ht]
\includegraphics[angle=90,width=\columnwidth]{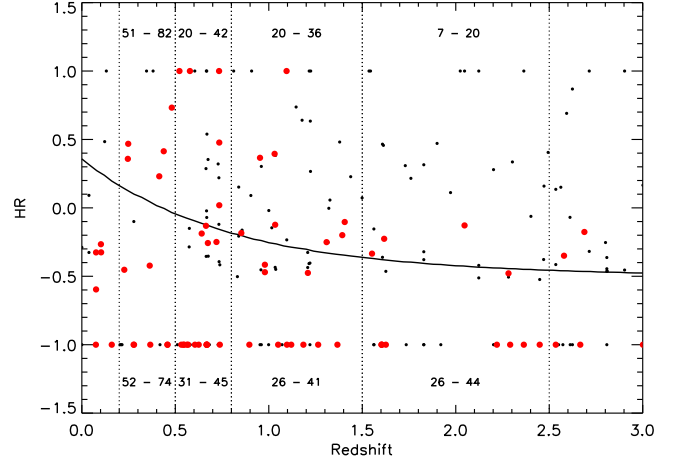}
\caption{The observed hardness ratio (HR) of GOODS-S X-ray AGNs as a function
of redshift.  Large red points show PACS detected AGNs and small
black points are PACS-undetected AGNs. Objects detected only in the hard(soft) 
bands have $HR=1$ and $HR=-1$ respectively. The solid black line shows the expected 
HR for a power-law X-ray spectrum with a typical photon index of $\Gamma=1.9$ 
and an obscuring Hydrogen column density of $N_H = 10^{22}$ cm$^{-2}$.
Dotted vertical lines mark the 4 redshift bins used in this study. The numbers in each
bin give the 1$\sigma$ range in PACS detection fraction (in \%) for obscured
X-ray AGNs (i.e, those above the solid line -- upper set of numbers) and unobscured
X-ray AGNs (i.e, those below the solid line -- lower set of numbers). A quick examination
will reveal that PACS detection fractions of obscured and unobscured AGNs are similar,
with only insignificant differences in any given redshift bin.
}
\label{nh_detfracs}
\end{figure}

In Fig.~\ref{nh_detfracs}, we plot $HR$ against redshift for AGNs in the GOODS-S field. Despite the lower
number statistics, we use this field for this plot because the PACS data is the deepest here and the detection fractions
are not swayed by the statistics of the most FIR luminous systems, typically strong starbursts.
The PACS detection fractions, estimated using Bayesian binomial statistics \citep{cameron11}, 
of AGNs above and below our fiducial boundary between unobscured and obscured
AGNs (solid curve: $N_H = 10^{22}$ cm$^{-2}$) are comparable in all four of our redshift bins.
Similar results are found in the GOODS-N and COSMOS fields: any differences in the detection
fractions between obscured and unobscured AGNs are at a 1$\sigma$ level and not significant,
given the typical uncertainties in X-ray counts. 

\section{Discussion}

Using deep X-ray catalogs and high-quality FIR imaging from Herschel/PACS, 
we have estimated the mean FIR luminosities
of AGNs across a wide span in redshift and SMBH accretion luminosity. We have studied the variation
of the 60 \mics\ specific luminosity with \lagn\ and find that \lfir\ is independent of  
\lagn\ at low accretion luminosities, but can show a strong correlation with \lagn\ at high accretion
luminosities. The strength of the correlation depends on redshift -- while strong at low redshifts ($z<0.8$), the correlation
appears to weaken or disappear at high redshifts. In this section, we discuss the implications of our results 
on our knowledge of the process of AGN fueling, its evolution with redshift and its dependence on the nature of the host galaxy.

The rest-frame FIR is widely regarded as the preferred tracer of star-formation in galaxies since it is
particularly sensitive to cool dust emission which dominates the output of star-forming regions.
However, as we show in \S3, the tail of FIR emission from hot dust heated by an active nucleus can
influence the FIR emission in AGN host galaxies, and in some cases, may dominate the 
60 \mics\ luminosity of such systems. While the likelihood of AGN dominance in the FIR depends positively
on the luminosity of the AGN, this is by no means a simple one-to-one relationship. Galaxies that
are luminous in the FIR (PACS detected AGNs) are usually SF-dominated with a small fraction that are
AGN-dominated. FIR weak galaxies (undetected in the PACS imaging) are, on average, 
SF-dominated as well. 
 
\subsection{Two regimes of AGN-galaxy co-evolution?}

Our results show that, among $z<1$ galaxies with relatively luminous AGNs (\lagn$\gtrsim 10^{44}$ \ergs), 
the mean nuclear activity closely tracks the bulk of star-formation.
At lower AGN luminosities, on the other hand, star-formation is unrelated to AGN luminosity. 
This suggests two regimes in the relationship between global star-formation in AGN hosts and nuclear activity.

The first regime, applicable to low luminosity AGNs, requires a disconnect between instantaneous
AGN accretion and host star-formation across the galaxy. The processes that drive and regulate most of the
star formation in such host galaxies do not directly fuel the SMBH. 
The second regime, applicable to high luminosity AGNs, 
is one in which the SMBH accretion tracks the total SFR of the host
galaxy. Here we expect direct coordination between global SF processes and
the fueling of the AGN, over timescales that are short compared to the duty cycle of AGN activity
\citep[$\sim 10^{7\rm{-}8}$ yr,][]{martini01, hopkins05a}. 

One of the best candidates for a process that directly links global SF and SMBH accretion 
are galaxy mergers, signatures of which are relatively common among local AGNs \citep{koss10}. 
Morphological studies of low-redshift star-forming QSOs also find frequent merger signatures 
\citep{canalizo01, urrutia08, veilleux09}. QSO hosts frequently
show bluer colors than typical old red ellipticals, in support of their recent merger origin \citep{jahnke04}.
However, a substantial fraction of QSO hosts show characteristics of undisturbed disks,
inconsistent with a pure merger origin for all QSOs \citep{bahcall97,dunlop03,guyon06}.
The question of whether the most luminous AGNs are always associated 
with mergers is still somewhat open. Part of the reason for this uncertainty is that the structural 
identification of bona-fide mergers is not entirely straightforward \citep[][and references therein]{lotz11}.
Studies of
moderate luminosity X-ray AGNs find no excess of merger signatures
compared to inactive galaxies \citep[e.g.][]{grogin05}, 
and only weak trends with AGN luminosity \citep{cisternas11, kocevski11}.
However, these studies are usually limited to AGN luminosities below \lagnc, with very limited
statistics at higher luminosities. In addition, our results seem to suggest that the role of mergers
as drivers of luminous AGNs may diminish at high redshifts, exactly where such
studies have concentrated. 

\subsection{The connection between SFR and \lagn\ in recent merger simulations}

\begin{figure}[t]
\includegraphics[width=\columnwidth]{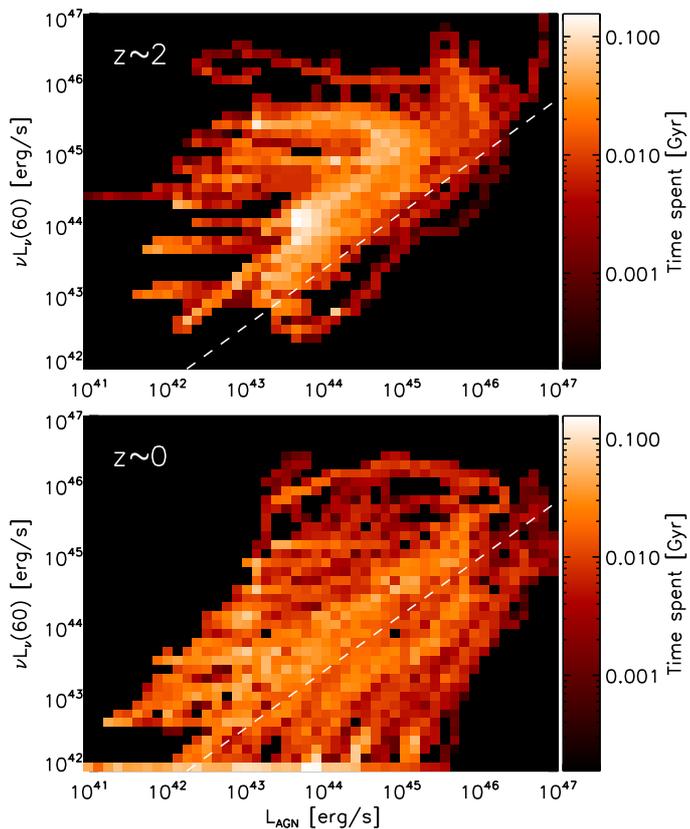}
\caption{Average time spent by hydrodynamic merger models (see text) in 
different regions of the plane defined by AGN luminosity and star forming 
luminosity. 
Top: z$\sim$2 mergers of gas rich progenitors. Bottom: local mergers.  
The dashed line indicates the relation 
$\nu L_\nu (60\mu m)\propto L_{AGN}^{0.8}$
of Netzer et al. (2009).
}
\label{merger_models}
\end{figure}

It is instructive
to test if the correlation observed among AGNs at $z\lesssim 1$,  
which we attribute to galaxy mergers, is indeed in agreement 
with models that trace the evolution of both star formation rate and 
accretion onto the black hole during a galaxy merger.

We have approached this by re-visualizing results from a set of  
current merger models. We use the models studied by 
\cite{wuyts10} which focus on progenitor galaxies
with size and high gas content as expected for z$\sim$2 mergers,
as well as a fully analogous set of models which encompass the larger sizes
and lower gas content of local mergers \citep{cox06}. In  
brief, all these models use the Gadget-2 smoothed particle hydrodynamic 
code \citep{springel05} and prescriptions for star formation, supernova
feedback, black hole growth and AGN feedback to trace the evolution 
of merging galaxies. For more details, we refer the reader to \cite{wuyts10} and references therein.  At
z$\sim$2, 25 simulations cover a number of interaction geometries of compact
high-z disks with gas fractions around 80\%, while at z$\sim$0 30 simulations
span different geometries for gas fractions 5--20\%.

We caution that this suite of models samples only a small part of the 
possible parameter space of progenitor properties and interaction geometries, 
and that the treatment of star formation and accretion is approximate -- the 
accretion rate follows the Bondi mechanism for the circumnclear conditions
at the model's spatial resolution of order 100~pc. 
However, these models represent among the best attempts at this time
to integrate SMBH accretion into galaxy mergers and can certainly  
allow a feasibility test of whether mergers produce an overall 
correlation between galaxy-wide SFR and \lagn.

The two panels of Fig.~\ref{merger_models} visualize the time that the mergers 
spend in different regions of the SFR/\lagn\ plane. These figures include tracks from
all merger models together, separated only by their redshift regime. Any individual  
merger model generally takes a rather complicated path in the SFR/\lagn\ plane, 
suggesting that there is strictly no real
tight synchronisation between the two processes. However, when taking the suite of 
models as a whole, it is clear that much time is indeed spent near the `correlation line'
from \cite{netzer09} and \S4.1, mostly when star formation and accretion decay 
after the merger. Such behavior is seen among mergers appropriate for both low
and high redshift progenitors.

We refrain from making further detailed conclusions due to the limitations of this approach
and since the scope of this discussion is limited to a statement of consistency. 
While much future modelling work remains to be done to improve the  tracing of mergers in
the SFR/\lagn\ plane, results from these current models are consistent  
with mergers spending considerable time in the general area of the 
`correlation line' of Fig.~\ref{shao_plot_main}. 

\subsection{Connections between global star-formation and AGN activity}

The existence of two regimes in the relationships between host SF and AGN activity
suggests that there exists one or more key processes which modulate both
star-formation and accretion onto the
nuclear black hole. Since the two regimes are distinguished by AGN luminosity,
this implies that the main mode of AGN fueling changes between low and high luminosity
AGNs. By this, we mean that the principal process (or family of processes) that drive 
the growth of AGNs is fundamentally different between the two regimes in AGN luminosity, 
and these two sorts of processes differ in their connection to global SF. 

Some insight into the processes that govern fueling at lower AGN luminosities
comes from the well-studied population of local Seyfert galaxies, which account for 
most of the low and moderate luminosity AGNs in the local Universe \citep[e.g.,][]{simkin80}. 
Seyfert host galaxies tend to be massive disks with substantial
spheroids, though some fraction are also in more elliptical hosts \citep{kauffmann03, kauffmann07, schawinski10}. 
While on-going major mergers are found among Seyferts, they account for only a minority of hosts. The statistics
of galaxy morphology and color among local Seyferts strongly disfavors a major merger
origin for the majority of local AGNs, since the big, stable disks in such galaxies
could not have undergone a recent merger event.

A plausible scenario, supported by such studies of the structure
and environment of local Seyfert galaxies is that SMBH fueling in low-luminosity AGNs
is driven mostly by stochastic infall of gas from the circumnuclear region of the host galaxy and bears
little connection to the extended SFR in the host \citep{kauffmann09}. Circumnuclear gas is replenished by several `secular'
processes that operate on the many galaxy-wide dynamical timescales.
The large differences between active periods of 
low-luminosity accretion and such long replenishment times
disconnects the nature of gas in the galaxy at large and the more rapid.

We find that the hosts of low luminosity AGNs are slightly, but measurably, different
from inactive galaxies. \cite{santini11} find that they have a slightly enhanced \emph{mean} SFR
compared to normal galaxies of the same mass and in Fig.~\ref{cosmos_colordist_all}, we 
show that AGN hosts have warmer than typical FIR colors. This implies that, while
the SFR in these hosts is not explicitly connected to AGN fueling, there is a preference
for AGNs to lie in galaxies that are currently forming stars.
The small shift to warmer colors among AGNs could be explained if AGNs are preferentially in star-forming
galaxies that have more central or circumnuclear SF, where ambient gas densities and dust
temperatures may be a bit higher than in the outskirts of galaxies. 
It may be that the processes that trigger bursts of SF in the centers of galaxies may also be related to gas
inflow to the nucleus, though not in a synchronised manner. This is supported
by high-resolution simulations of gas inflow to the inner few parsecs in disk galaxies \citep{hopkins10}.

Luminous AGNs (\lagn$>$\lagnc), on the other hand, are fueled by rapid nuclear inflow of gas associated with 
major mergers of gas-rich galaxies \citep{menci03, volonteri03, hopkins06,  hopkins08a, somerville08}. 
Since they also inspire strong star-bursts, major mergers are a good candidate for a galaxy-wide 
process that directly correlates AGN accretion and host star-formation.
In a sense, the merger-driven starburst-QSO connection may be thought of as a scaled up version of the 
SF--accretion relationship that may more commonly found among low-luminosity AGN,  
since in mergers the total SFR of the host galaxy is dominated by the nuclear starburst. 

\subsection{The role of major mergers at high redshift}

Turning now to $z\gtrsim1$, our results show that the correlation between SF and AGN luminosity
weakens or disappears at these redshifts. We proceed with an interpretation
taking the points as they are measured, but caution that the interpretation is subject to the
uncertainties associated with the measurement of \lfir\ for the most luminous AGNs.

Guided by our explanations in the above discussion of low redshift trends, we
can develop a couple of explanations for the flattening of the high redshift relationship.
It may be that  the secular processes 
active in low-redshift Seyferts are capable
of fueling most high luminosity AGNs as well. Several studies have established that
gas-rich, turbulent disks are common among massive galaxies
at high redshift, due to high inflow rates of cold gas
\citep{erb06, forsterschreiber06, forsterschreiber09}. In such disks, a greater turbulent viscosity leads
to a higher SMBH accretion rate \citep{bournaud11, bournaud11aph}, and possibly
more frequent high Eddington-ratio bursts,
without necessarily producing a direct connection between accretion rate and SFR.
If greater turbulent accretion can fuel luminous AGNs, the role of mergers
could be eclipsed at high redshift by fueling through such secular means.

Having said this, current simulations still do not have the resolution to link large stochastic
bursts of accretion to enhanced secular inflow, so the idea developed above is not
strongly motivated by theory. 
If, instead, luminous AGNs at all redshifts are attributed to major mergers,
then the lack of a correlation at high redshifts could occur
if a regime of SF-AGN synchronisation, associated locally
with the final coalescence of major gas-rich mergers, is lost among distant galaxies. 
This may happen if, for e.g, the interval of time between major mergers at high
redshifts is short compared to the timescale of peak accretion, as cosmological
simulations seem to suggest.

\subsection{Implications for black hole-galaxy coevolution}

If the correlation between \lfir\ and \lagn\ is indeed due to a dominant role of mergers among
luminous AGNs, as supported by studies of local and low redshift populations, our results
suggest a change in the relevance or nature of major galaxy mergers as drivers of
AGN-galaxy co-evolution at high redshifts. We turn now to the implications of this in the
context of galaxy evolution.

The most popular explanation for the form and tightness of the well-known local
M$_{BH}$-$\sigma$ relationship is that most black hole
growth occurs in major mergers, where both the galaxy and the SMBH grow in lock-step.
These high luminosity, high accretion-rate phases of SMBH growth are visible
in their end stages as QSOs and evolutionary models of SMBH growth by mergers
can explain simultaneously the form and evolution of 
empirical SMBH mass functions, QSO luminosity functions, merger and red galaxy density
functions, clustering, and the X-ray background \citep{hopkins08a, hopkins08b, somerville08}.

Our results suggest that at low-z, 
most massive BH growth occurs in mergers or other possible processes that subject
the host to violent relaxation. This maintains and tightens the massive end of local
SMBH scaling relations. 

At high-z, mergers may be less important. Many studies show that high-z galaxies are quite different
from local ones. SF galaxies typically have large gas fractions and high turbulent line-widths 
\citep{genzel08,forsterschreiber09, daddi10, tacconi10}.
Such galaxies can have a high duty-cycle of unstable modes (clumps, bars, grand spirals)
which can strongly torque gas. In addition, the larger magnitude of turbulent viscosity compared
to local settled gas disks also plausibly leads to greater  net `secular' inflow to the nucleus.  
Simulations suggest that sustained inflow rates are enhanced at these redshifts \citep{bournaud11}.

It is unclear if this larger net accretion rate to the nuclear regions translates into an increase
in the frequency of low-luminosity AGNs, as suggested by \cite{bournaud11},
or a boost across the board in the density of AGNs at all luminosities. The flattening of the faint-end slope with 
redshift from X-ray luminosity functions \citep{silverman08, aird10} supports the latter explanation. If so,
the flat \lfir-\lagn\ trend at $z>1$ suggests that the role of mergers in fueling high luminosity AGNs
may be replaced by global disk instabilities in gas-rich galaxies, which effectively removes the synchronisation
between AGN accretion rate and SFR inherent to major mergers. Secular processes, 
which are important for low-luminosity AGNs at low redshifts, become dominant at all AGN luminosities at $z>1$.

Regardless of the mechanism, our results suggest that the lock-step growth of stellar mass and 
galaxy properties becomes effectively decoupled from the growth of the black hole at higher redshifts. 
Most models of SMBH-galaxy co-evolution require that black holes grow in concert with their host
galaxy (at least its spheroidal component). However, at the redshifts where both the stellar content
and black holes in massive galaxies are put into place, the correlation between SMBH and galaxy
growth seems to weaken or disappear. This suggests that alternative ideas are needed to tie
BH growth to galaxy growth, in order to yield the tight  M$_{BH}$-$\sigma$ relation seen today.
Possible ideas are a strong role for feedback-regulated BH growth, coupled with a short lag between
BH and bulge growth \citep{hopkins09}, or perhaps pure statistical evolution of both components
in a hierarchical merging context \citep{jahnke11}.

\section{Summary}

We have explored the variation of the mean far-IR luminosity of X-ray selected active galaxies
with AGN luminosity and obscuration, using deep X-ray and Herschel/PACS datasets
in three key extragalactic survey fields: GOODS-S, GOODS-N and COSMOS.
Through the combination of both narrow+deep and wide+shallow surveys,
redshift binning and the inclusion of stacked measurements for FIR-undetected 
subsamples, we avoid the effects of Malmquist bias which can produce 
incorrect correlations between SFR and AGN properties in
FIR detected samples taken only from wide+shallow surveys.

We gauge the importance of hot dust emission from the active nucleus at FIR wavelengths, using
a suite of empirical AGN SEDs and SEDs of star-forming galaxies, chosen to cover most of the
range of SED shapes expected for these two galaxy components. We compare the observed-frame
MIR and FIR colors of AGNs and IR-bright inactive (X-ray undetected) galaxies in bins of
redshift, finding that active galaxies generally span roughly the same color range as inactive galaxies
at FIR wavelengths, though with a small tendency to host warmer FIR SEDs. We also demonstrate
that a small minority of AGNs, typically the most luminous ones, are dominated by hot dust
emission from the active nucleus.

After appropriately combining measurements from all three fields, we find characteristic trends between 
AGN luminosity and FIR luminosity \lfir. Low-luminosity AGNs display essentially no relationship between global
star-formation rate, as traced by FIR luminosity, and the luminosity of the active nucleus.
Instead, the \lfir\ of such AGNs increases with redshift as expected from studies of the
SFR evolution of inactive massive galaxies. Black hole fueling in these systems is probably 
dominated by secular processes related to gas inflow and momentum transfer in galaxy disks 
and is unrelated to the overall growth of the host. 

On the other hand, luminous AGNs (\lagn$>10^{44.8}$ \ergs) at low and moderate redshift
($z<1$) show a strong correlation between SFR and nuclear luminosity, implying a close
relationship between the growth of the stellar component of the host galaxy and the growth
of the SMBH.
We interpret this as evidence for the importance of gas-rich major mergers 
in fueling high accretion phases in AGNs at these redshifts, 
consistent with the results of modern hydrodynamic simulations.

At higher redshifts ($z>1$), we find that the correlation flattens or disappears, implying
a weaker connection between galaxy and black hole growth. This suggests that
that the role of major mergers as drivers of black hole-galaxy co-evolution evolves with 
redshift. If so, this has important implications for co-evolutionary models and the evolution of the
M$_{BH}$-$\sigma$ relationship.

We find very little dependence between the mean FIR luminosity and X-ray obscuring
column $N_H$ in X-ray selected AGNs, over five orders of magnitude in $N_H$.
The population of obscured AGNs do not distinguish themselves
in terms of their current SFR and various mechanisms are 
may be responsible for obscuring the nuclear X-ray emission of AGNs, 
with some certainly unrelated to global star-formation in the host galaxy.

\begin{acknowledgements}
We thank Marta Volonteri for valuable discussions.
PACS has been developed by a consortium of institutes led by MPE (Germany) and including UVIE
(Austria); KU Leuven, CSL, IMEC (Belgium); CEA, LAM (France); MPIA (Germany); INAF-IFSI/
OAA/OAP/OAT, LENS, SISSA (Italy); IAC (Spain). This development has been supported by the
funding agencies BMVIT (Austria), ESA-PRODEX (Belgium), CEA/CNES (France), DLR (Germany),
ASI/INAF (Italy), and CICYT/MCYT (Spain). FB acknowledges support from Financiamento
Basal, CONICYT-Chile FONDECYT 1101024 and FONDAP-CATA 15010003, and
Chandra X-ray Center grant SAO SP1-12007B (F.E.B.)
\end{acknowledgements}

\bibliographystyle{aa}

\bibliography{pep_agn_sfr}

\appendix

\section{FIR luminosities: detection, stacking and measurement}

As a direct tracer of the FIR emission we concentrate on the mean luminosity \lums, estimated at a rest-frame wavelength of 60 \mics. This choice is a compromise between a wavelength long enough to avoid most of the AGN contamination (see discussion below) and at the same time short enough to be sampled by PACS 160 \mics\ observations even at the highest redshifts considered in this work. 

To preclude any assumptions about the SED shape, we computed \lums\ though a log-linear interpolation of PACS fluxes, after converting them to luminosities following the approach of \cite{shao10}. For the GOODS-S dataset, which includes 70 \mics\ observations as well, we interpolated between the two PACS bands bracketing rest-frame 60 \mics. 
We checked that wider wavelength  coverage in GOODS-S (the addition of 70 \mics\ observations) does not introduce any bias in the analysis: when interpolating rest-frame 60 \mics\ luminosity only from 100 and 160 \mics\ PACS photometry, our results remain unchanged, and individual \lums\ do not differ by more than 50\%. 

We also explored fitting PACS fluxes with a typical IR template (e.g. from \citealt{ce01} or \citealt{dh02} libraries) to derive \lums. 
We found differences by a few tens per cent (only in 15\% of the bins is the difference larger than $50\%$), 
but the global picture is unaffected. Therefore, we adopt the simple interpolated \lums\ estimate throughout, making no assumptions on the detailed SED shape. 

For sources fully detected (i.e. detected in both PACS bands used to interpolate 60 \mics\ luminosity) we used the corresponding fluxes and computed individual $\nu L^i_\nu (60\mu $m$)$. 
For PACS sources that were completely undetected in both bands, we computed average  fluxes by stacking in a given \lhard\ and redshift interval. We stacked at the X-ray positions on PACS residual maps using the \cite{bethermin10} libraries. The use of residual maps, from which all detected objects were removed, avoids contamination by nearby brighter sources. PSF photometry was performed on the final stacked images. 
For each PACS band $j$, we then averaged these stacked fluxes with individual fluxes of partially detected objects (i.e. detected only in band $j$ but not in the other one), weighting by the number of sources. 
These stacked and averaged fluxes in each band are used to get $\nu L^{STACK}_\nu (60\mu $m$)$, in the same fashion as for the fully detected sources.

The final 60 \mics\ luminosity in each \lhard\ and redshift interval is computed by averaging over the linear luminosities of detections and non-detections, weighted by the number of sources. Only bins with more than 3 sources are used in our analysis.

Errors on the infrared luminosity were obtained by bootstrapping, in similar fashion as many current studies \citep{shao10, mullaney11b,santini11}. A set of N sources, where N is equal to the number of sources per bin in \lhard\ and redshift, is randomly chosen 100 times among detections and non-detections (allowing repetitions), and a \lums\ is computed per each iteration. The standard deviation of the obtained \lums\ values gives the error on the average 60 \mics\ luminosity in each bin. The error bars thus account for both measurement errors and the error on the population distribution. They do not, however, account for cosmic variance. 

\section{Combination across fields}

\begin{figure*}[ht]
\includegraphics[width=\textwidth]{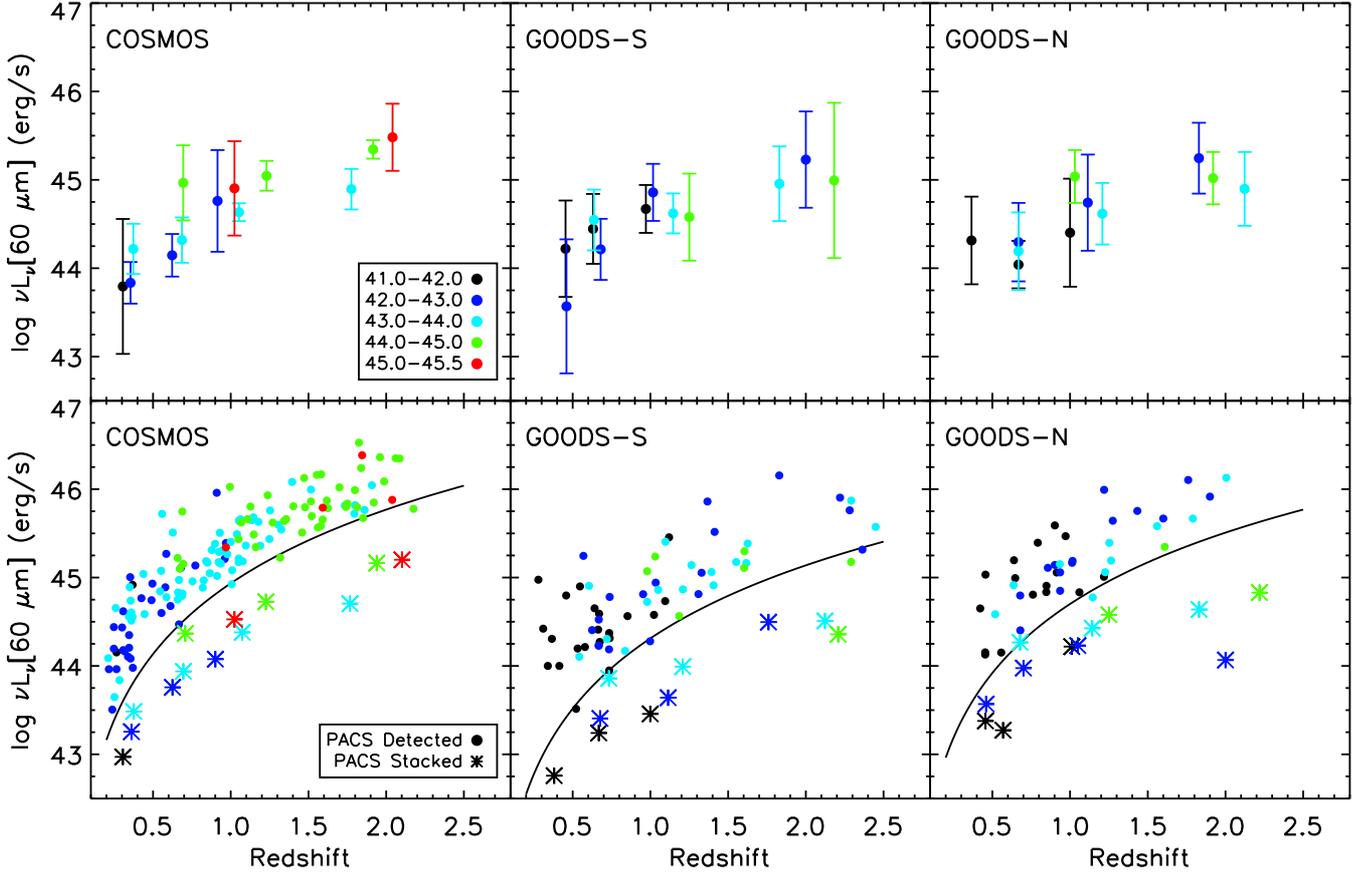}
\caption{
\lums\ as a function of redshift for X-ray AGNs in the individual COSMOS, GOODS-S and GOODS-N fields.
The points in all panels are colored by logarithmic bins in hard band X-ray luminosity (\lhard), as shown.
The upper panels show the mean \lfir\ estimated by combining fluxes from PACS detections and stacks.
Note the consistency across all fields between the values and variation with redshift of the mean measurements.
The lower panels show individual PACS detected AGNs and the stacked points for PACS undetected AGNs
(in bins of redshift and \lhard). The solid black lines show the approximate $3\sigma$ \lums\
limit, derived from the flux limits of the 100 and 160 \mics\ PEP photometry catalogs.
}
\label{z_vs_l60}
\end{figure*}

To reduce scatter and the effects of cosmic variance, we combine the estimates of  
the average \lums\ in each bin in redshift and X-ray luminosity from 
all three fields into one mean number for each bin. 

The PACS maps in each of the fields differ in depth and area. Therefore,
we begin by comparing the estimates of \lums\ in all three fields against redshift, as a simple test to ensure
that the mean measurements in all three fields are in fact compatible. In the three upper panels of
Fig.~\ref{z_vs_l60}, we plot the mean \lums\ from AGNs in each of the three fields separately.
The points are colored by bins in X-ray luminosity. At a glance, one may gather that both 
the measurements and their redshift evolution in all three fields are very comparable. 

In the lower three panels of Fig.~\ref{z_vs_l60}, we plot separately the \lums\ of PACS detected AGNs
and those derived from stacks of the undetected AGNs. In addition, we have plotted the approximate
rest-frame 60 \mics\ luminosity limit as a function of redshift, derived from a log-linear 
interpolation of the 3$\sigma$ PACS detection limits at 100 and 160 \mics\ for each field. These luminosity limits
match reasonably well the lower envelope of the PACS detected AGNs in the Figure. The COSMOS 
detections are more luminous than those in the GOODS fields both at the luminosity limit, due to the significantly
shallower PACS imaging, as well as at the upper end, because the larger area of the COSMOS field
($5\times$ larger than the GOODS fields combined) brings in more of the rare, very FIR luminous
galaxies into the sample. The stacked points can be up to an order of magnitude
fainter than the PACS luminosity limit. 

Due to the differences in the depth and noise properties of the PACS images in the three different fields, 
combining the data at the level of the maps poses difficulties. 
On the other hand, the measurements of FIR luminosity in each of the three fields show consistent
values and behavior with redshift and \lhard. Therefore we
combined {\it post-facto} the estimates of the mean \lums\ in each redshift and \lhard\ bin. 
This was done by taking an average of the FIR luminosity in all three fields in each bin with a valid measurement, 
weighted by the inverse variance (1/err$^2$) derived from the errors on the mean FIR luminosity.

\end{document}